\documentclass[english,aps,preprint,superscriptaddress]{revtex4-1}
\usepackage[T1]{fontenc}
\usepackage[latin9]{inputenc}
\usepackage{array}
\usepackage{multirow}
\usepackage{amsmath}
\usepackage{graphicx}
\usepackage{esint}

\makeatletter

\providecommand{\tabularnewline}{\\}

\@ifundefined{textcolor}{}
{%
 \definecolor{BLACK}{gray}{0}
 \definecolor{WHITE}{gray}{1}
 \definecolor{RED}{rgb}{1,0,0}
 \definecolor{GREEN}{rgb}{0,1,0}
 \definecolor{BLUE}{rgb}{0,0,1}
 \definecolor{CYAN}{cmyk}{1,0,0,0}
 \definecolor{MAGENTA}{cmyk}{0,1,0,0}
 \definecolor{YELLOW}{cmyk}{0,0,1,0}
 }

\makeatother

\usepackage{babel}
\begin{document}

\preprint{This line only printed with preprint option}

\title{Revised self-consistent continuum solvation in electronic-structure
calculations}

\author{Oliviero Andreussi}

\email{oliviero.andreussi@epfl.ch}

\selectlanguage{english}%

\affiliation{Department of Materials Science and Engineering, Massachusetts Institute
of Technology, Cambridge, MA 02139, USA}

\affiliation{Theory and Simulations of Materials, École Polytechnique Fédérale
de Lausanne, Station 12, 1015 Lausanne, Switzerland}

\author{Ismaila Dabo}

\email{daboi@cermics.enpc.fr}

\selectlanguage{english}%

\affiliation{CERMICS, Project-team INRIA Micmac, Université Paris-Est, 77455 Marne-la-Vallée,
France}

\author{Nicola Marzari}

\email{nicola.marzari@epfl.ch}

\selectlanguage{english}%

\affiliation{Theory and Simulations of Materials, École Polytechnique Fédérale
de Lausanne, Station 12, 1015 Lausanne, Switzerland}

\begin{abstract}
The solvation model proposed by Fattebert and Gygi \cite{fattebert_jcomputchem_2002}
and Scherlis et al. \cite{scherlis_jcp_2006} is reformulated, overcoming
some of the numerical limitations encountered and extending its range
of applicability. We first recast the problem in terms of induced
polarization charges that act as a direct mapping of the self-consistent
continuum dielectric; this allows to define a functional form for
the dielectric that is well behaved both in the high-density region
of the nuclear charges and in the low-density region where the electronic
wavefunctions decay into the solvent. Second, we outline an iterative
procedure to solve the Poisson equation for the quantum fragment embedded
in the solvent that does not require multi-grid algorithms, is trivially
parallel, and can be applied to any Bravais crystallographic system.
Last, we capture some of the non-electrostatic or cavitation terms
via a combined use of the quantum volume and quantum surface \cite{cococcioni_prl_2005}
of the solute. The resulting self-consistent continuum solvation (SCCS)
model provides a very effective and compact fit of computational and
experimental data, whereby the static dielectric constant of the solvent
and one parameter allow to fit the electrostatic energy provided by
the PCM model with a mean absolute error of 0.3 kcal/mol on a set
of 240 neutral solutes. Two parameters allow to fit experimental solvation
energies on the same set with a mean absolute error of 1.3 kcal/mol.
A detailed analysis of these results, broken down along different
classes of chemical compounds, shows that several classes of organic
compounds display very high accuracy, with solvation energies in error
of 0.3-0.4 kcal/mol, whereby larger discrepancies are mostly limited
to self-dissociating species and strong hydrogen-bond forming compounds.
\end{abstract}
\maketitle

\section{introduction}

Continuum solvation models have proved to be very effective in capturing
the complexity of a solvent in an implicit fashion \cite{tomasi_chemrev_1994,cramer_chemrev_1999,orozco_chemrev_2000,tomasi_chemrev_2005}.
Even if one could argue that an explicit description of the solvent
would be a more faithful representation of the real system, an explicit
solvent would require extensive molecular dynamics (MD) simulations
in order to obtain meaningful thermodynamic averages. In addition,
even if these were computationally feasible, an ab-initio density-functional
theory (DFT) description of the solvent would rarely provide the correct
results, due to the limitations in the accuracy of the functionals
adopted. Critical limitations in the representation of the structure
of liquid water with DFT are well known in the literature \cite{laasonen_jcp_1993,sprik_jcp_1996,grossman_jcp_2004,vandevondele_jcp_2005,sit_jcp_2005},
and can be related to the lack of a proper description of van der
Waals interactions and hydrogen bonds in standard DFT. Moreover, given
the presence of hydrogen atoms in liquid water, neglecting the quantum
motion of the nuclei of the system could significantly affect the
computed physical properties of the system \cite{stern_jcp_2001,chen_prl_2003,morrone_prl_2008}.
The above limitations in the DFT description of liquid water translate
into phase diagrams that differ from the experimental one, and melting
temperatures that are much larger than the experimental temperature
\cite{sit_jcp_2005,vandevondele_jcp_2005}. Being unable to accurately
describe the structure and the dynamics of liquid water, it is questionable
whether the most popular DFT functionals could correctly reproduce
its dielectric behavior. As a typical dipolar liquid, the dielectric
properties of liquid water derive from the electric dipole moment
carried by the individual molecules. Even though much effort has been
put in the characterization of the dipole moment of water in its liquid
state, a consensus on the subject is still lacking \cite{silvestrelli_prl_1999,pasquarello_prb_2003,sharma_prl_2007}.
Moreover, it was shown in the literature \cite{sharma_prl_2007} that
the dielectric response of water is dominated by the short range effects
of the hydrogen bond environment, thus implying that the lack of accuracy
in the description of intermolecular interactions in liquid water
could lead to a poor characterization of its dielectric properties. 

Many continuum solvation models have been proposed and widely developed
\cite{tomasi_chemrev_1994,cramer_chemrev_1999,orozco_chemrev_2000,tomasi_chemrev_2005},
especially in the chemistry literature, since the earliest work of
Onsager \cite{onsager_jacs_1936}. Within these many approaches, one
of the most popular is the Polarizable Continuum Model (PCM) of Tomasi
and coworkers \cite{tomasi_chemrev_1994,tomasi_chemrev_2005} that,
in its latest formulation in terms of Integral Equations (IEF) \cite{mennucci_jpcb_1997,cances_jmathchem_1998},
encompasses a wide range of similar methods (e.g. the COSMO approach
\cite{klamt_jchemsoc_1993}). Being mostly linked to the chemistry
community, PCM has not been used in condensed matter and solid state
simulations. In particular, the possibility to deal with metallic
systems within PCM was introduced only later in an implicit way \cite{corni_jcp_2001},
and the algorithm was not really developed to be interfaced with periodic
systems and ab-initio MD simulations. 

In an effort to extend solvation methods to plane-wave, periodic-boundary
codes and ab-initio MD, Fattebert and Gygi proposed a new model of
continuum solvation \cite{fattebert_jcomputchem_2002,fattebert_intjqchem_2003},
where the dielectric is defined as a smooth self-consistent function
of the electronic density of the solute. This model was further extended
by Scherlis et al. \cite{scherlis_jcp_2006} to include the calculation
of the cavitation energy, by defining it in terms of the quantum surface
of the solute \cite{cococcioni_prl_2005}. Despite the simple and
elegant formulation of the model, some ill-conditioning of the problem
in its original formulation led to abandoning or relaxing self-consistency
\cite{dziedzic_epl_2011}: models in which the dielectric is defined
in terms of a fictitious, atom-centered electronic density have been
proposed \cite{sanchez_jcp_2009}, together with approaches were the
electronic density of the solute precomputed in vacuum is adopted
to define the continuum solvent. In addition, all of the reported
implementations and derivations of the Fattebert and Gygi method (e.g.
Refs. \cite{scherlis_jcp_2006,dziedzic_epl_2011}) rely on multigrid
solvers, which entail high-order discretization of the Laplacian operator
and that typically require Cartesian geometries and serial implementations. 

In the present work, starting from the model of Fattebert and Gygi,
a revised self-consistent continuum solvation (SCCS) model is derived,
in which the challenges outlined above are tackled as follows. First,
along the lines of PCM \cite{tomasi_chemrev_2005}, the method is
reimplemented in terms of polarization charges and solved using an
iterative approach that is intrinsically parallel, extendable to any
kind of Bravais lattice, and straightforwardly transferable into any
plane-wave code. Second, some of the numerical instabilities of the
original formulation \cite{fattebert_jcomputchem_2002,fattebert_intjqchem_2003}
are solved by properly redefining the relation between the dielectric
and the electronic density of the solute. Third, the model features
a revised version of the extension of Scherlis et al. \cite{scherlis_jcp_2006,cococcioni_prl_2005}
to treat the cavitation contribution to solvation free energies, where
the concepts of quantum surfaces and quantum volumes \cite{cococcioni_prl_2005}
have been extended to model dispersion and repulsion effects in a
simplified way. 

The accuracy of the proposed method has been tested on a reference
sample of 240 neutral solutes in water. Its ability to accurately
reproduce with only one fitting parameter PCM results provides a strong
validation of the method and allows its extension to reproduce experimental
solvation energies. The resulting agreement with experiments has a
mean absolute error of 1.3 kcal/mol when two fitting parameters are
used. Moreover, an analysis of the error as a function of the functional
groups of the solutes provides some critical insights into potential
limitations of the model. In particular, results for several classes
of organic compounds present very high accuracies (with errors on
the order of 0.3-0.4 kcal/mol), and discrepancies are mostly limited
to chemical compounds that undergo dissociation in water, such as
carboxylic acids and amines. Significant errors are also found for
some strong hydrogen-bond forming species, such as ethers, alcohols
and fluorinated compounds. These findings suggest that an extension
of the model that takes into account self-dissociation or a shell
of explicit water molecules around the solute could provide results
in quantitative agreement with experiment. 

The paper is organized as follows. In Section \ref{sec:Previous-approaches}
the main features of previous continuum solvation methods are reviewed.
In Section \ref{sec:Method} the basic equations of the electrostatic
part of the proposed method are presented. In Section \ref{sec:Choice-of-dielectric}
some limitations of previous models are discussed and a new definition
of the dielectric function that solves most of these limitations is
presented. In Section \ref{sec:Iterative-vs-Multigrid} a new numerical
approach is discussed, based on an iterative solution of the dielectric
problem, alternative of and simpler than multigrid solvers. In Section
\ref{sec:Extra-terms}, complementary non-electrostatic contributions
to solvation are introduced. Eventually, in Section \ref{sec:Results}
parametrizations of the model and a comparison with theoretical results
from similar well assessed techniques are reported.

\section{\label{sec:Previous-approaches}Previous approaches}

The most widespread continuum solvation method is the Polarizable
Continuum Model by Tomasi and coworkers \cite{tomasi_chemrev_2005},
that in its formulation in terms of integral equations \cite{mennucci_jpcb_1997,cances_jmathchem_1998}
represents a most general approach to continuum solvation. The basic
physical picture behind the model is one of a solute contained in
an ad-hoc cavity surrounded by a continuous polarizable dielectric,
whose response to the solute charge distribution is fully characterized
by the value of its static dielectric constant $\epsilon_{0}$. In
this model, the transition between the vacuum region inside the solute
cavity and the surrounding dielectric continuum is sharp and discontinuous.
This discontinuity allows to treat the effect of the surrounding environment
on the solute through introducing a polarization charge density that
is exactly localized at the vacuum-dielectric interface. In addition
to homogeneous isotropic dielectrics, a suitable surface charge density
can be defined via integral equations also in complex embedding environments,
that include multiple interfaces and metallic systems. Thus, by representing
the response of the environment in terms of an effective surface polarization
density, IEF-PCM reduces a three-dimensional problem into a two-dimensional
one. Numerically, IEF-PCM adopts a Boundary Element Method (BEM) to
discretize the surface of the molecular cavity and the operators defined
on the surface domain \cite{pascualahuir_jcomputchem_1990,silla_jmolgraph_1990}.
The final ingredient is the definition of the molecular cavity: in
this respect, different choices have been adopted \cite{scalmani_jcp_2010},
the most popular one being a rigid cavity built as the superposition
of atom-centered spheres with fixed radii, corresponding to empirical
van der Waals atomic radii multiplied by a solvent-dependent scaling
factor. This choice allows to have a regular discretization of the
cavity surface that helps numerical convergence (see Refs.\cite{scalmani_jcp_2010,lipparini_jcp_2010}
and references therein). On the other hand, the numerical discretization
of the cavity surfaces has lead, in the original formulations, to
atomic forces that were not continuous with respect to atomic positions
and suffered from numerical singularities. For this reason it has
been difficult to extend PCM to ab-initio MD simulations. In order
to solve such a problem, modified versions of continuum models have
been proposed in the literature\cite{senn_jcp_2003,lange_jcp_2010}.
Moreover, it is worth noting that more advanced definitions of the
molecular cavity have been proposed in terms of an isodensity of the
electronic density of the solute. Isosurfaces of both the frozen electronic
density of the solute (Isodensity PCM \cite{foresman_jpc_1996}) and
the self-consistent density (SCI-PCM \cite{wiberg_jpc_1995}) have
been considered. Nonetheless, few applications of the above methods
have been reported in the literature, probably because of the absence
of analytic gradients. 

In order to extend continuum solvation to ab-initio MD simulations,
a novel approach has been proposed by Fattebert and Gygi \cite{fattebert_jcomputchem_2002,fattebert_intjqchem_2003}.
The main difference with respect to PCM is in the definition of a
dielectric function $\epsilon\left(\mathbf{r}\right)\equiv\epsilon\left(\rho^{elec}\left(\mathbf{r}\right)\right)$,
defined in terms of the electronic density $\rho^{elec}$ of the solute,
and smoothly varying between 1, when $\rho^{elec}$ is large, to $\epsilon_{0}$
when $\rho^{elec}\rightarrow0$. The physical problem is expressed
in terms of the electrostatic response generated by the embedding
dielectric and acting on the solute. This response field, which is
defined in the whole three-dimensional space, is then obtained numerically
by using a multigrid solver. The method of Fattebert and Gygi requires
- as it is argued here - a careful choice for the dielectric function
in terms of the electronic density. Moreover, its implementation in
plane-wave (PW) ab-initio codes has been hindered by the necessity
to interface such codes with an efficient, high-order, ideally parallel
multigrid solver able to work in the arbitrary geometry of any Bravais
lattice.

\section{\label{sec:Method}Present model}

We show now that, starting from the basic equations of Fattebert and
Gygi \cite{fattebert_jcomputchem_2002,fattebert_intjqchem_2003},
it is possible to recast the electrostatic problem in terms of a polarization
density, similarly to what is done in PCM. The key assumption is that
of a dielectric medium self-consistently modeled on the electronic
density of the solute, via a suitably defined relationship
\begin{equation}
\epsilon\left(\mathbf{r}\right)=\epsilon\left(\rho^{elec}\left(\mathbf{r}\right)\right),\label{eq:general_dielectric}
\end{equation}
such that the dielectric is excluded ($\epsilon=1$) from the inner
part of the solute, where the electronic density is high, while it
smoothly goes to the bulk dielectric constant of the solvent ($\epsilon=\epsilon_{0}$)
outside the solute, where the electronic density goes to zero. By
adding a dielectric medium to the system, the electrostatic field
that enters into the quantum-mechanical problem is no longer given
by the Poisson equation in vacuum 
\begin{equation}
\nabla^{2}\phi^{tot}\left(\mathbf{r}\right)=-4\pi\rho^{solute}\left(\mathbf{r}\right),\label{eq:poisson_vacuum}
\end{equation}
where the total charge density in vacuum is the sum of the electronic
and ionic densities 
\begin{equation}
\rho^{solute}\left(\mathbf{r}\right)=\rho^{elec}\left(\mathbf{r}\right)+\rho^{ions}\left(\mathbf{r}\right),\label{eq:solute_density}
\end{equation}
but has to be the solution of the more complex Poisson equation
\begin{equation}
\mathbf{\nabla}\cdot\epsilon\left(\rho^{elec}\left(\mathbf{r}\right)\right)\mathbf{\nabla}\phi^{tot}\left(\mathbf{r}\right)=-4\pi\rho^{solute}\left(\mathbf{r}\right)\label{eq:poisson_solvent}
\end{equation}
that is nothing but the standard Maxwell equation 
\begin{equation}
\nabla\cdot\mathbf{D}\left(\mathbf{r}\right)=4\pi\rho^{solute}\left(\mathbf{r}\right),\label{eq:maxwell_displacement}
\end{equation}
for the displacement field 
\begin{equation}
\mathbf{D}\left(\mathbf{r}\right)=\mathbf{E}\left(\mathbf{r}\right)+4\pi\mathbf{P}\left(\mathbf{r}\right)=\epsilon\left(\mathbf{r}\right)\mathbf{E}\left(\mathbf{r}\right),\label{eq:maxwell_definitiondisplacement}
\end{equation}
where $\mathbf{E}\left(\mathbf{r}\right)$ and $\mathbf{P}\left(\mathbf{r}\right)$
are the electric field and the polarization.

Eq. (\ref{eq:maxwell_displacement}) can be transformed into an equation
in terms of the electric field
\begin{equation}
\nabla\cdot\mathbf{E}\left(\mathbf{r}\right)=4\pi\rho^{solute}\left(\mathbf{r}\right)-4\pi\nabla\cdot\mathbf{P}\left(\mathbf{r}\right)\label{eq:maxwell_electricfield}
\end{equation}
and, by mapping the effect of the dielectric into a polarization charge
density
\begin{equation}
\rho^{pol}\left(\mathbf{r}\right)\equiv-\nabla\cdot\mathbf{P}\left(\mathbf{r}\right)=\mathbf{\nabla}\cdot\left(\frac{\epsilon\left(\rho^{elec}\left(\mathbf{r}\right)\right)-1}{4\pi}\mathbf{\nabla}\phi^{tot}\left(\mathbf{r}\right)\right),\label{eq:rhopol_definition}
\end{equation}
a vacuum-like Poisson problem is recovered 
\begin{equation}
\nabla^{2}\phi^{tot}\left(\mathbf{r}\right)=-4\pi\left(\rho^{solute}\left(\mathbf{r}\right)+\rho^{pol}\left(\mathbf{r}\right)\right).\label{eq:poisson_rhopol}
\end{equation}
As in PCM, the nonlinear nature of the problem that arises from the
mutual polarization of the solvent and solute is accounted for via
a polarization charge density that depends on itself through the total
electrostatic potential. 

By evaluating the gradient on the right hand side of Eq. (\ref{eq:rhopol_definition})
and performing some simple algebraic manipulations, the polarization
charge density can be expressed alternatively as the sum of two distinct
terms:
\begin{equation}
\rho^{pol}\left(\mathbf{r}\right)=\frac{1}{4\pi}\mathbf{\nabla}\ln\epsilon\left(\rho^{elec}\left(\mathbf{r}\right)\right)\cdot\mathbf{\nabla}\phi^{tot}\left(\mathbf{r}\right)-\frac{\epsilon\left(\rho^{elec}\left(\mathbf{r}\right)\right)-1}{\epsilon\left(\rho^{elec}\left(\mathbf{r}\right)\right)}\rho^{solute}\left(\mathbf{r}\right).\label{eq:rhopol_final}
\end{equation}
 As in the case of PCM, introducing polarization charges allows to
deal with localized physical quantities. Indeed, both terms in the
above equations are only defined in a narrow region around the solute.
In particular, the first term contains all the nonlinear character
of the problem, and is confined within the vacuum-solvent interface,
where the dielectric function has a gradient different from zero.
The second term, instead, is a simple rescaling of the solute charge
density that extends into the vacuum region (see Ref. \cite{chipman_jcp_2006}
and references therein): although in principle defined in the whole
three-dimensional space, this term is also localized around the solute
owing to the exponential decay of the electronic charge density in
the dielectric region. 

It should be noted that formulating the problem in terms of the polarization
charge density instead of the solvent electrostatic potential allows
- thanks to the Gauss theorem - to define a sum rule for the total
polarization charge surrounding the solute, namely,
\begin{equation}
\int\rho^{pol}\left(\mathbf{r}\right)d\mathbf{r}=-\frac{\epsilon_{0}-1}{\epsilon_{0}}\int\rho^{solute}\left(\mathbf{r}\right)d\mathbf{r}.\label{eq:sum_rule}
\end{equation}
Eventually, from the linearity of the Poisson equation Eq. (\ref{eq:poisson_rhopol}),
it follows that the total field can also be written as 
\begin{equation}
\phi^{tot}\left(\mathbf{r}\right)=\phi^{solute}\left(\mathbf{r}\right)+\phi^{pol}\left(\mathbf{r}\right).\label{eq:total_field}
\end{equation}
where the $\phi^{solute}$ or $\phi^{pol}$ fields are solutions of
vacuum-like Poisson equations in terms of the $\rho^{solute}$ or
$\rho^{pol}$ charge densities. Such a formal separation of the potential
is useful in order to express all of the important quantities that
enter into quantum-mechanical simulations (total energy, potentials,
forces, etc.) in terms of two contributions: one explicitly depending
on the solute charge density alone (identified, now and in the following,
by the superscript $solute$) and one explicitly depending on the
dielectric (identified by the superscript $pol$). The first term
is analogous to a quantity computed in vacuum and it is readily provided
by the simulation code adopted. On the other hand, the polarization
contributions need to be explicitly added, in order to include the
effect of the dielectric in the simulation. Although the method can
be applied to different quantum-mechanical approaches, in the present
article we will limit our discussion to the case of Density Functional
Theory (DFT) in the Kohn-Sham formulation. The derivation of the polarization
contributions to the electrostatic energy, Kohn-Sham energy, Kohn-Sham
potential, and atomic forces is presented below.

\subsubsection{Electrostatic Energy}

Once the Poisson equation is solved and the polarization of the dielectric
is known, the electrostatic energy of the system can be expressed
as 
\begin{equation}
E^{el}=\frac{1}{8\pi}\int\mathbf{E}\left(\mathbf{r}\right)\cdot\mathbf{D}\left(\mathbf{r}\right)d\mathbf{r}=\frac{1}{8\pi}\int\epsilon\left(\rho^{elec}\left(\mathbf{r}\right)\right)\left|\nabla\phi^{tot}\left(\mathbf{r}\right)\right|^{2}d\mathbf{r}.\label{eq:energy_solvent_fg}
\end{equation}
By integrating by parts the last expression, the electrostatic energy
can be expressed as 
\begin{equation}
E^{el}=\frac{1}{2}\int\rho^{solute}\left(\mathbf{r}\right)\phi^{tot}\left(\mathbf{r}\right)d\mathbf{r}.\label{eq:energy_solvent_fg_integrated}
\end{equation}
The same expression can be obtained starting from the vacuum-like
problem in Eq. (\ref{eq:poisson_rhopol}), by expressing the electrostatic
energy as the sum of the total interaction energy of the charge densities
of the system, both the solute and the polarization one, 
\begin{equation}
E^{int}=\frac{1}{2}\int\rho^{tot}\left(\mathbf{r}\right)\phi^{tot}\left(\mathbf{r}\right)d\mathbf{r}=\frac{1}{2}\int\left(\rho^{solute}\left(\mathbf{r}\right)+\rho^{pol}\left(\mathbf{r}\right)\right)\phi^{tot}\left(\mathbf{r}\right)d\mathbf{r}\label{eq:total_interaction_energy}
\end{equation}
with the addition of the work done to polarize the dielectric. Such
a work, assuming a linear behavior for the dielectric, can be shown
to be 
\begin{equation}
W=-\frac{1}{2}\int\mathbf{P}\left(\mathbf{r}\right)\cdot\mathbf{E\left(\mathbf{r}\right)}d\mathbf{r}=-\frac{1}{2}\int\rho^{pol}\left(\mathbf{r}\right)\phi^{tot}\left(\mathbf{r}\right)d\mathbf{r}\label{eq:work_to_polarize}
\end{equation}
that, combined with Eq.(\ref{eq:total_interaction_energy}), correctly
provides the result in Eq. (\ref{eq:energy_solvent_fg_integrated}).

Adopting the decomposition of the potential introduced in Eq. (\ref{eq:total_field}),
the electrostatic energy of the system can be further written as
\begin{equation}
E^{el}=E^{solute}+E^{pol}\label{eq:energy_el}
\end{equation}
where the first term is the electrostatic energy of the solute, including
both electrons and ions, 
\begin{equation}
E^{solute}=\frac{1}{2}\int\rho^{solute}\left(\mathbf{r}\right)\phi^{solute}\left(\mathbf{r}\right)d\mathbf{r},\label{eq:energy_vacuum}
\end{equation}
and is the analogous of the electrostatic energy of the system in
vacuum. Now, by exploiting the fact that 
\begin{equation}
\frac{1}{2}\int\rho^{solute}\left(\mathbf{r}\right)\phi^{pol}\left(\mathbf{r}\right)d\mathbf{r}=\frac{1}{2}\int\int\frac{\rho^{solute}\left(\mathbf{r}\right)\rho^{pol}\left(\mathbf{r'}\right)}{\left|\mathbf{r}-\mathbf{r'}\right|}d\mathbf{r}d\mathbf{r'}=\frac{1}{2}\int\rho^{pol}\left(\mathbf{r}\right)\phi^{solute}\left(\mathbf{r}\right)d\mathbf{r},\label{eq:equivalence_solute_polarization}
\end{equation}
the polarization term can be included in two equivalent ways: following
PCM, as the integral of the polarization charge density times the
potential of the solute charge densities acting on it, or, viceversa,
as the integral of the solute charge density times the polarization
potential acting on it 
\begin{eqnarray}
E^{pol} & = & \frac{1}{2}\int\rho^{pol}\left(\mathbf{r}\right)\phi^{solute}\left(\mathbf{r}\right)d\mathbf{r}\label{eq:energy_solvent1}\\
 & = & \frac{1}{2}\int\rho^{solute}\left(\mathbf{r}\right)\phi^{pol}\left(\mathbf{r}\right)d\mathbf{r}.\label{eq:energy_solvent2}
\end{eqnarray}

\subsubsection{Kohn-Sham Energy and Potential}

From the above equations, the total DFT energy of the system can be
expressed as 
\begin{multline}
E^{tot}\left[\rho^{elec},\rho^{ions}\right]=E^{kin}\left[\rho^{elec}\right]+E^{el}\left[\rho^{elec},\rho^{ions}\right]+E^{xc}\left[\rho^{elec}\right]\\
=E^{kin}\left[\rho^{elec}\right]+E^{solute}\left[\rho^{elec},\rho^{ions}\right]+E^{xc}\left[\rho^{elec}\right]+E^{pol}\left[\rho^{elec},\rho^{ions}\right]\\
=\left(E^{tot}\left[\rho^{elec},\rho^{ions}\right]\right)_{solute}+E^{pol}\left[\rho^{elec},\rho^{ions}\right]\label{eq:kohn-sham_energy}
\end{multline}
where, in the last equality, we underline the fact that the quantity
\begin{equation}
\left(E^{tot}\left[\rho^{elec},\rho^{ions}\right]\right)_{solute}\equiv E^{kin}\left[\rho^{elec}\right]+E^{solute}\left[\rho^{elec},\rho^{ions}\right]+E^{xc}\left[\rho^{elec}\right]\label{eq:kohn-sham_solute}
\end{equation}
does not depend explicitly on the polarization charge distribution
and, thus, is analogous to the total Kohn-Sham energy of a system
in vacuum. 

The effective DFT potential acting on the electrons can be written
as (see, for example, Appendix A of Ref \cite{sanchez_jcp_2009}),
\begin{equation}
\frac{\delta E^{tot}\left[\rho^{elec},\rho^{ions}\right]}{\delta\rho^{elec}}=\left(\frac{\delta E^{tot}\left[\rho^{elec},\rho^{ions}\right]}{\delta\rho^{elec}}\right)_{solute}+v_{pol}+v_{\epsilon},\label{eq:total_functional}
\end{equation}
where the first term on the right-hand side includes all of the contributions
to the potential that do not depend explicitly on the polarization
charge density and is analogous to the potential of a system in vacuum.
As for the two additional terms, the first one is a purely local potential
contribution and is simply given by the electrostatic potential generated
by the polarization charge density 
\begin{equation}
v_{pol}\left(\mathbf{r}\right)=\phi^{pol}\left(\mathbf{r}\right),\label{eq:v_polarization}
\end{equation}
while the second term arises from the dependence of the dielectric
on the electronic charge, as defined in Eq. (\ref{eq:general_dielectric}),
and is given by
\begin{equation}
v_{\epsilon}\left(\mathbf{r}\right)=-\frac{1}{8\pi}\frac{d\epsilon\left(\rho^{elec}\left(\mathbf{r}\right)\right)}{d\rho^{elec}}\left|\mathbf{\nabla}\phi^{tot}\left(\mathbf{r}\right)\right|^{2}.\label{eq:v_epsilon}
\end{equation}

\subsubsection{Forces}

Exploiting the Helmann-Feynman theorem, the force $\mathbf{f_{a}}$
acting on an atom $a$ of the solute is given by 
\begin{equation}
\mathbf{f}_{a}^{tot}=-\frac{\partial E^{tot}}{\partial\mathbf{R}_{a}}=-\frac{\partial E^{el}}{\partial\mathbf{R}_{a}}\label{eq:force_zero}
\end{equation}
where $\mathbf{R}{}_{a}$ is the position of atom $a$ and we exploit
the fact that the electronic kinetic energy and exchange-correlation
potentials do not depend explicitly on nuclear positions. By using
the expression for the electrostatic energy in Eq. (\ref{eq:energy_solvent_fg}),
the force is given by
\begin{multline}
\mathbf{f}_{a}^{tot}=-\frac{\partial}{\partial\mathbf{R}_{a}}\frac{1}{8\pi}\int\epsilon\left(\rho^{elec}\left(\mathbf{r}\right)\right)\left|\nabla\phi^{tot}\left(\mathbf{r}\right)\right|^{2}d\mathbf{r}\\
=-\frac{1}{8\pi}\left[\int\left|\nabla\phi^{tot}\left(\mathbf{r}\right)\right|^{2}\frac{\partial\epsilon\left(\rho^{elec}\left(\mathbf{r}\right)\right)}{\partial\mathbf{R}_{a}}d\mathbf{r}+\int\epsilon\left(\rho^{elec}\left(\mathbf{r}\right)\right)\frac{\partial}{\partial\mathbf{R}_{a}}\left|\nabla\phi^{tot}\left(\mathbf{r}\right)\right|^{2}d\mathbf{r}\right]\label{eq:force_fg}
\end{multline}
Provided that the dielectric permittivity is not an explicit function
of the atomic positions of the solute, as is the case of the definition
adopted in Eq. (\ref{eq:general_dielectric}), the first contribution
in the above equation is equal to zero. The second contribution, instead,
can be expressed as
\begin{multline}
\mathbf{f}_{a}^{tot}=-\frac{1}{8\pi}\int\epsilon\left(\rho^{elec}\left(\mathbf{r}\right)\right)\frac{\partial}{\partial\mathbf{R}_{a}}\left|\nabla\phi^{tot}\left(\mathbf{r}\right)\right|^{2}d\mathbf{r}\\
=-\frac{1}{4\pi}\int\epsilon\left(\rho^{elec}\left(\mathbf{r}\right)\right)\nabla\phi^{tot}\left(\mathbf{r}\right)\cdot\frac{\partial}{\partial\mathbf{R}_{a}}\nabla\phi^{tot}\left(\mathbf{r}\right)d\mathbf{r}\\
=-\frac{1}{4\pi}\int\epsilon\left(\rho^{elec}\left(\mathbf{r}\right)\right)\nabla\phi^{tot}\left(\mathbf{r}\right)\cdot\nabla\frac{\partial}{\partial\mathbf{R}_{a}}\phi^{tot}\left(\mathbf{r}\right)d\mathbf{r}\\
=-\int\left[-\frac{1}{4\pi}\nabla\cdot\epsilon\left(\rho^{elec}\left(\mathbf{r}\right)\right)\nabla\phi^{tot}\left(\mathbf{r}\right)\right]\frac{\partial}{\partial\mathbf{R}_{a}}\phi^{tot}\left(\mathbf{r}\right)d\mathbf{r}\\
=-\int\rho^{solute}\frac{\partial}{\partial\mathbf{R}_{a}}\phi^{tot}\left(\mathbf{r}\right)d\mathbf{r}.\label{eq:force_fg_simplified}
\end{multline}
By applying the partial differential to the first formulation of the
electrostatic energy in Eq. (\ref{eq:energy_solvent_fg}), an alternative
equation for the force is obtained, namely 
\begin{multline}
\mathbf{f}_{a}^{tot}=-\frac{\partial}{\partial\mathbf{R}_{a}}\frac{1}{2}\int\rho^{solute}\left(\mathbf{r}\right)\phi^{tot}\left(\mathbf{r}\right)d\mathbf{r}\\
=-\frac{1}{2}\int\phi^{tot}\left(\mathbf{r}\right)\frac{\partial\rho^{solute}\left(\mathbf{r}\right)}{\partial\mathbf{R}_{a}}d\mathbf{r}-\frac{1}{2}\int\rho^{solute}\left(\mathbf{r}\right)\frac{\partial\phi^{tot}\left(\mathbf{r}\right)}{\partial\mathbf{R}_{a}}d\mathbf{r},\label{eq:force_simple}
\end{multline}
which, combined with Eq. (\ref{eq:force_fg_simplified}), provides
the useful relation
\begin{equation}
\int\rho^{solute}\left(\mathbf{r}\right)\frac{\partial\phi^{tot}\left(\mathbf{r}\right)}{\partial\mathbf{R}_{a}}d\mathbf{r}=\int\phi^{tot}\left(\mathbf{r}\right)\frac{\partial\rho^{solute}\left(\mathbf{r}\right)}{\partial\mathbf{R}_{a}}d\mathbf{r}.\label{eq:equivalence_forces}
\end{equation}
Similarly to the expressions for the electrostatic and the Kohn-Sham
energies in Eqs. (\ref{eq:energy_el}) and (\ref{eq:kohn-sham_energy}),
also the interatomic forces can be defined as sums of two terms
\begin{align}
\mathbf{f}_{a}^{tot}=- & \int\rho^{solute}\left(\mathbf{r}\right)\frac{\partial}{\partial\mathbf{R}_{a}}\phi^{solute}\left(\mathbf{r}\right)d\mathbf{r}-\int\rho^{solute}\left(\mathbf{r}\right)\frac{\partial}{\partial\mathbf{R}_{a}}\phi^{pol}\left(\mathbf{r}\right)d\mathbf{r}\label{eq:force_two_terms}\\
= & \left(\mathbf{f}_{a}^{tot}\right)_{solute}+\mathbf{f}_{a}^{pol}.
\end{align}
The first contribution does not depend explicitly on the polarization
charge density and is analogous to the interatomic force on a system
in vacuum. By using Eq. (\ref{eq:equivalence_forces}) and similarly
to what is done in Eq. (\ref{eq:equivalence_solute_polarization}),
the polarization contribution to interatomic forces can also be expressed
as 
\begin{align}
\mathbf{f}_{a}^{pol}= & -\int\phi^{pol}\left(\mathbf{r}\right)\frac{\partial}{\partial\mathbf{R}_{a}}\rho^{solute}\left(\mathbf{r}\right)d\mathbf{r}\label{eq:force_pol1}\\
= & -\int\rho^{pol}\left(\mathbf{r}\right)\frac{\partial}{\partial\mathbf{R}_{a}}\phi^{solute}\left(\mathbf{r}\right)d\mathbf{r},\label{eq:force_pol2}
\end{align}
where the partial derivative is now applied to analytic functions
of nuclear positions. 

When the dielectric depends instead explicitly on atomic positions,
Eq. (\ref{eq:equivalence_forces}) is no longer exact and additional
contributions to forces would arise. This is the case, for example,
when the dielectric constant is entirely defined in terms of a fictitious
electronic density centered at the atomic positions, as described
in Ref. \cite{sanchez_jcp_2009}. It is important to notice that similar
contributions to the forces arise also when the dielectric is defined
in terms of the sum of the electronic density plus a fictitious ionic
density, as is the case in, e.g., the original Fattebert and Gygi
model, where additional core charges were added to saturate the dielectric
constant in the solute region \cite{fattebert_jcomputchem_2002,fattebert_intjqchem_2003}.
In this case, a contribution similar to the one derived in Ref. \cite{sanchez_jcp_2009}
should be explicitly added to the forces, unless the derivative of
the dielectric with respect to this fictitious nuclear density is
zero, i.e., unless this density is added only in a region of space
where the resulting dielectric constant is flat.

\section{\label{sec:Choice-of-dielectric}Choice of dielectric function}

Although formally equivalent to the original Gygi-Fattebert model,
the equations presented above have the advantage of highlighting the
numerical challenges of the original formulation. In particular, the
main ingredient of the model is the dielectric function that appears
in Eq. (\ref{eq:general_dielectric}), and for the model to work properly
and seamlessly some conditions on this function should be imposed.
These conditions are as follows:

\begin{figure}
\includegraphics[width=0.8\textwidth]{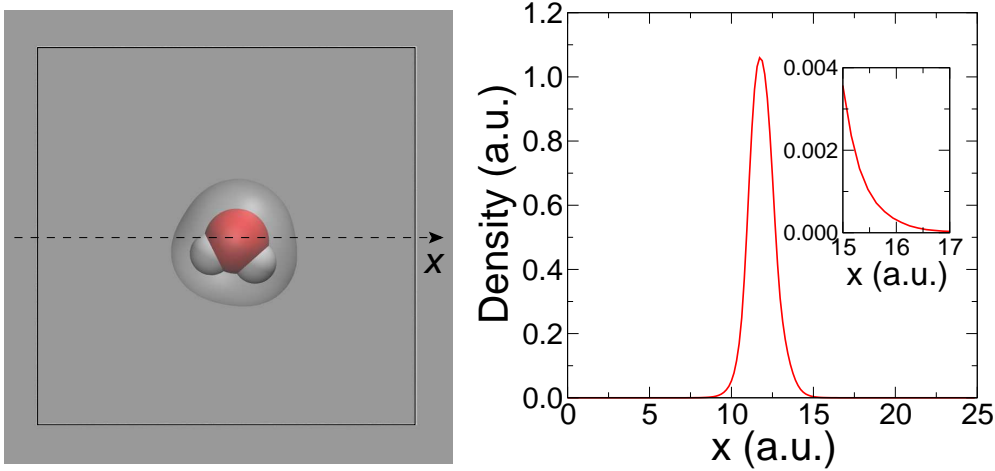}\caption{\label{fig:Test-simulation}Numerical test. Left: single water molecule
in a periodic cubic cell of $\left(25\textrm{ a.u.}\right){}^{3}$
size. The isosurface of the density corresponding to a threshold value
of 0.0003 a.u. is shown. Right: electronic density of the water molecule
along the x axis passing through the center of the oxygen atom. }
\end{figure}

\begin{enumerate}
\item Extrema: the dielectric function should go monotonically from a value
of 1 (vacuum) inside the molecule to a value of $\epsilon_{0}$ in
the bulk of the solvent:
\begin{equation}
\begin{cases}
\lim_{\rho^{elec}\rightarrow0}\epsilon\left(\rho^{elec}\right) & =\epsilon_{0}\\
\lim_{\rho^{elec}\rightarrow\infty}\epsilon\left(\rho^{elec}\right) & =1
\end{cases}\label{eq:eps_extrema}
\end{equation}
 
\item Flatness inside the solute: the dielectric should be exactly equal
to one above a certain density threshold, to avoid spurious polarization
effects due to the interaction of the dielectric with the ion cores. 
\item Flatness in bulk solvent: the dielectric should be exactly equal to
$\epsilon_{0}$ below a certain threshold, to avoid spurious polarization
charges in the bulk of the solvent, due to potential numerical noise
in the exponentially vanishing electronic density away from the solute.
\item Smoothness: since we are interested in implementing the model in a
plane-wave, periodic electronic-structure code, the dielectric function
and its gradient have to be smooth enough to be well described in
a three-dimensional grid that has a resolution given by the typical
density cutoffs used in plane-wave DFT calculations. Such condition
is crucial to the convergence of the iterative SCF calculation. Moreover,
since our formulation of the method relies on the polarization density
of the medium, this later quantity should also be smooth enough to
be well described with the density cutoffs that are conventionally
adopted.
\end{enumerate}
The original dielectric function by Fattebert and Gygi \cite{fattebert_jcomputchem_2002,fattebert_intjqchem_2003}
\begin{equation}
\epsilon_{\epsilon_{0},\rho_{0},\beta}\left(\rho^{elec}\right)=1+\frac{\left(\epsilon_{0}-1\right)}{2}\left(1+\frac{1-\left(\rho^{elec}/\rho_{0}\right)^{2\beta}}{1+\left(\rho^{elec}/\rho_{0}\right)^{2\beta}}\right)\label{eq:eps_original}
\end{equation}
satisfies the first requirement, going very smoothly from $\epsilon=1$
in the proximity of the solute to $\epsilon=\epsilon_{0}$ in the
bulk of the solvent. 
\begin{figure}
\includegraphics[width=0.9\textwidth]{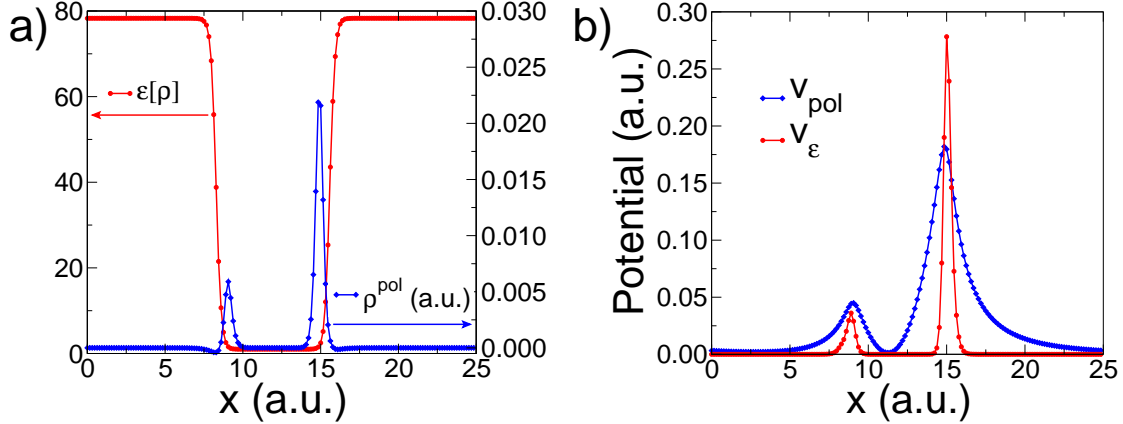}\caption{\label{fig:FG-quantities}a) Dielectric function (re) and polarization
charge (blue) of the water molecule of Figure (\ref{fig:Test-simulation})
plotted along the x axis. b) Polarization field {[}Eq. (\ref{eq:v_polarization}){]}
and additional self-consistent dielectric contribution to electronic
Hamiltonian {[}Eq. (\ref{eq:v_epsilon}){]} for the same system. Physical
quantities are computed using the original formulation by Fattebert
and Gygi {[}Eq. (\ref{eq:eps_original}){]} with the parameters reported
in Ref. \cite{scherlis_jcp_2006}. The filled circles correspond to
the points of the real space grid used in the calculation, corresponding
to a density cutoff of 400 Ry.}
\end{figure}
The smoothness of the function allows it to be well defined in the
three-dimensional mesh used in typical calculations (see Figure (\ref{fig:FG-quantities})).
Nonetheless, convergence issues may arise in some particular systems.
Such issues have been reported by Sanchez et al. \cite{sanchez_jcp_2009}
for simulations of two-dimensional systems (i.e. slabs), and result
from the fact that the gradient of the dielectric function and the
dielectric-dependent contribution to the potential in the energy functional
are too abruptly varying to be correctly described with typical grid
sizes. It is worth noting here that the extra potential term $v_{\epsilon}$
of Eq. (\ref{eq:v_epsilon}) is somewhat similar to the polarization
charge density: both quantities depends on the gradient of the dielectric
and of the total field, thus their are both confined in the small
region around the solute and they both suffer for the same conditioning
problems (compare, for example, the two quantities in Figure (\ref{fig:FG-quantities})).
Last, the smoothness of the Fattebert and Gygi function leads to a
polarization density in a region very close to the nuclei, potentially
breaking the second requirement. As mentioned in Section \ref{sec:Method},
this problem has already been pointed out in the original paper by
Fattebert and Gygi, and it has been solved by using an additional
density in the definition of $\epsilon\left(\rho\left(\mathbf{r}\right)\right)$,
in order to force the dielectric constant to go to $\epsilon=1$ in
the proximity of the nuclei \cite{fattebert_jcomputchem_2002,fattebert_intjqchem_2003}.
While this approach is correct in theory, in practice it introduces
some additional terms in the forces acting on the nuclei that have
to be explicitly considered (see final discussion in the previous
Section). Moreover, the final result of the calculation depends strongly
on the choice of the introduced additional density or on the choice
of parameters used to model the ionic density (i.e. shape of the nuclei,
Gaussian spread, etc.). For these reasons it appears that the original
function, at least in the parametrization adopted in Ref \cite{scherlis_jcp_2006},
is excessively smooth where it should not (close to the nuclei), while
not smooth enough where it should (outside the molecule).

With the aim of correcting these drawbacks, we decided to introduce
a new expression for the dielectric function by using a piecewise
definition of the dielectric of the form
\begin{equation}
\epsilon_{\epsilon_{0},\rho_{min},\rho_{max}}\left(\rho^{elec}\right)=\begin{cases}
1 & \rho^{elec}>\rho_{max}\\
s\left(\rho^{elec}\right) & \rho_{min}<\rho^{elec}<\rho_{max}\\
\epsilon_{0} & \rho^{elec}<\rho_{min}
\end{cases}\label{eq:eps_trial}
\end{equation}
where $s\left(x\right)$ is a general smooth switching function that
decreases monotonically from $s\left(\rho_{min}\right)=\epsilon_{0}$
to $s\left(\rho_{max}\right)=1$. Several switching functions $s\left(x\right)$
were considered, e.g. the trigonometric function 
\begin{equation}
s\left(x\right)=1+\frac{\epsilon_{0}-1}{2\pi}\left[2\pi\frac{\left(\rho_{max}-x\right)}{\left(\rho_{max}-\rho_{min}\right)}-\sin\left(2\pi\frac{\left(\rho_{max}-x\right)}{\left(\rho_{max}-\rho_{min}\right)}\right)\right].\label{eq:switch_trial}
\end{equation}
Nonetheless, despite their similarity with the original function of
Fattebert and Gygi, all of the trial functions resulted in even worse
or no convergence of the electronic-structure calculation at any reasonable
cutoffs. The reason for such a ill-conditioning is that the region
where the $s\left(\rho^{elec}\right)$ function is defined lies outside
of the molecule, where the electronic density decays exponentially.
Thus, the resulting function $s\left(\mathbf{r}\right)$ also decays
exponentially. Nevertheless, since the electronic density is known
to vanish exponentially, the problematic analytical behavior can be
rectified easily by redefining the switching function in terms of
the logarithm of the density, namely, 
\begin{equation}
s'\left(\rho^{elec}\right)=s\left(\ln\rho^{elec}\right)=1+\frac{\epsilon_{0}-1}{2\pi}\left[2\pi\frac{\left(\ln\rho_{max}-\ln\rho^{elec}\right)}{\left(\ln\rho_{max}-\ln\rho_{min}\right)}-\sin\left(2\pi\frac{\left(\ln\rho_{max}-\ln\rho^{elec}\right)}{\left(\ln\rho_{max}-\ln\rho_{min}\right)}\right)\right].\label{eq:switch_improved}
\end{equation}

In order to increase the smoothness of the dielectric function and
of the resulting polarization charges, a further condition can be
imposed on the form of the switching function. In particular, from
Eq. (\ref{eq:rhopol_final}) it is clear that the polarization charges
are mostly given by a term of the form $\mathbf{\nabla}\ln\epsilon\left(\rho^{elec}\left(\mathbf{r}\right)\right)\cdot\mathbf{\nabla}\phi^{total}\left(\mathbf{r}\right)$.
While no assumptions can be made, a priori, on the behavior of the
gradient of the total field, in defining the dielectric function it
is convenient to choose a form of the switching function such that
the derivative of its logarithm is well behaved. This can be easily
achieved by selecting a dielectric function of the form
\begin{equation}
\epsilon_{\epsilon_{0},\rho_{min},\rho_{max}}\left(\rho^{elec}\right)=\begin{cases}
1 & \rho^{elec}>\rho_{max}\\
\exp\left(t\left(\ln\rho^{elec}\right)\right) & \rho_{min}<\rho^{elec}<\rho_{max}\\
\epsilon_{0} & \rho^{elec}<\rho_{min}
\end{cases},\label{eq:eps_new}
\end{equation}
where $t\left(x\right)$ is a general, smooth function that decreases
monotonically from $t\left(\ln\rho_{min}\right)=\ln\epsilon_{0}$
to $t\left(\ln\rho_{max}\right)=0.$ 
\begin{figure}
\includegraphics[width=0.95\textwidth]{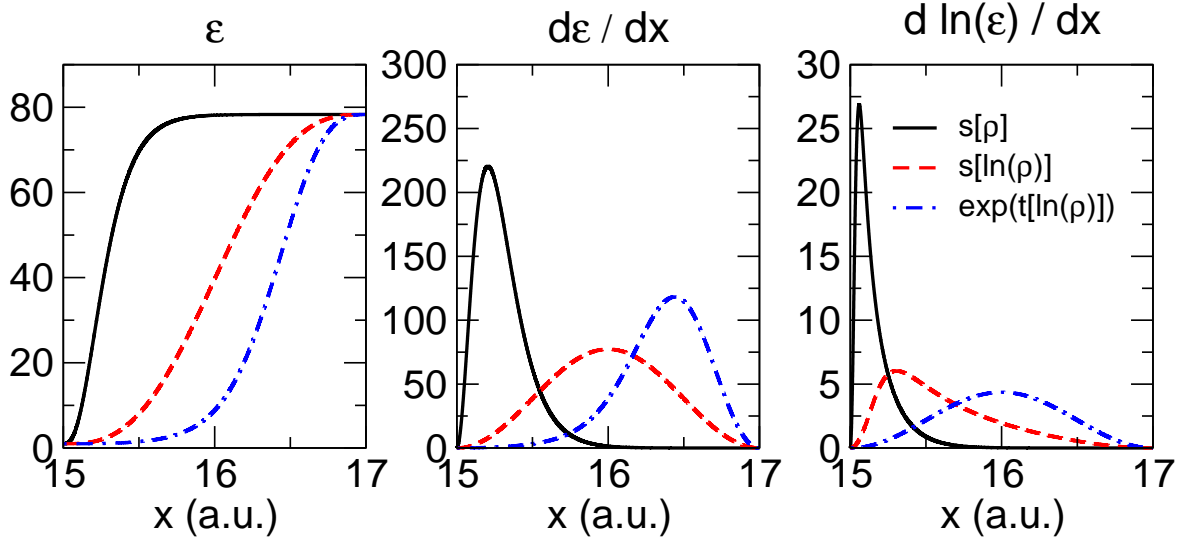}

\caption{Behaviors of the dielectric function (left), its derivative (center)
and the derivative of its logarithm (right) for the three different
switching functions reported in Eq. (\ref{eq:switch_trial}), black,
Eq. (\ref{eq:switch_improved}), red, and Eq. (\ref{eq:switch_new}),
blue.}
\end{figure}
Explicitly, in this work, we decided to adopt the trigonometric function
\begin{equation}
t\left(x\right)=\frac{\ln\epsilon_{0}}{2\pi}\left[2\pi\frac{\left(\ln\rho_{max}-x\right)}{\left(\ln\rho_{max}-\ln\rho_{min}\right)}-\sin\left(2\pi\frac{\left(\ln\rho_{max}-x\right)}{\left(\ln\rho_{max}-\ln\rho_{min}\right)}\right)\right],\label{eq:switch_new}
\end{equation}
whose first derivative 
\begin{equation}
\frac{dt\left(x\right)}{dx}=\frac{-\ln\epsilon_{0}}{\left(\ln\rho_{max}-\ln\rho_{min}\right)}\left[1-\cos\left(2\pi\frac{\left(\ln\rho_{max}-x\right)}{\left(\ln\rho_{max}-\ln\rho_{min}\right)}\right)\right],\label{eq:switch_derivative}
\end{equation}
vanishes with zero slope at the extrema of the interval of definition,
thus making $\epsilon$ and its first two derivatives continuous in
the whole space. It is important to notice that the proposed function
does not contain any internal adjustable parameters, but only depends
on the value of the dielectric constant and on the two density thresholds
$\rho_{min}$ and $\rho_{max}$.

\section{\label{sec:Iterative-vs-Multigrid}Iterative vs Multigrid}

In order to effectively compute the polarization field in Eq. (\ref{eq:poisson_solvent})
and related polarization charges, different numerical procedures can
be adopted. While most of the previous implementations rely on multigrid
solvers, we have found more advantageous to rely on an iterative procedure,
derived along the lines of a similar approach first introduced in
PCM.

\begin{figure}
\includegraphics[width=0.9\textwidth]{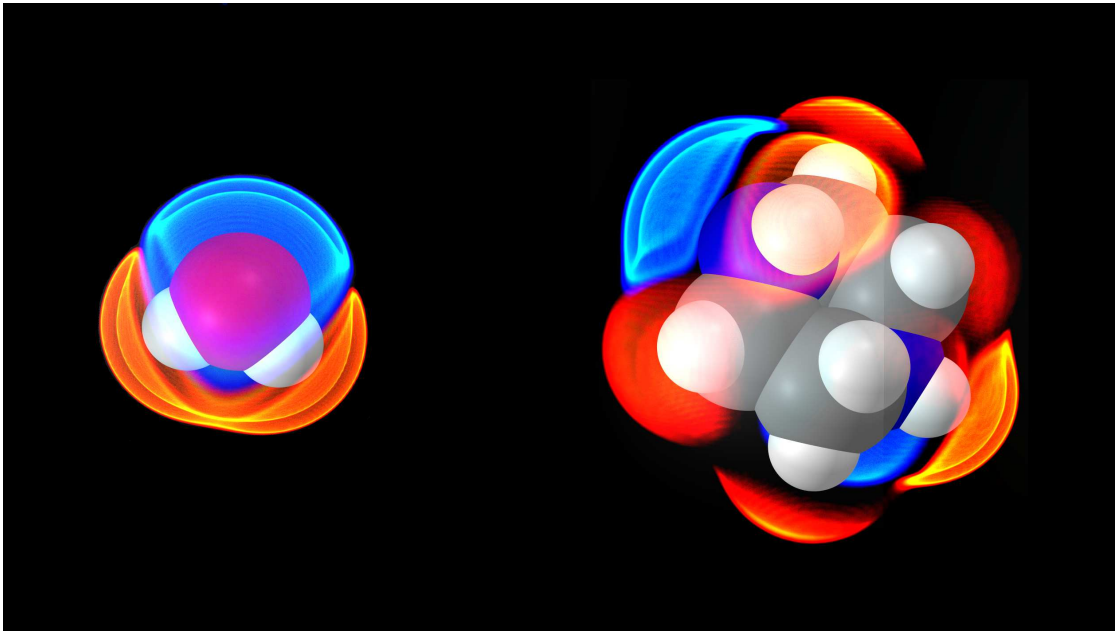}

\caption{\label{fig:h2o_c4h10n2}Positive (orange) and negative (blue) polarization
charges of a water molecule, on the left, and of a piperazine molecule,
on the right (molecule 013 of the training set, structure in Figure
(\ref{fig:Fitting-set})). Brighter (darker) colors correspond to
values of the density around $\pm$0.03 a.u. ($\pm$0.01 a.u.). Stronger
polarization charge densities are found around the lone pairs of electrons
of the etheroatoms (oxygen in red and nitrogen in blue) and on the
more polarized hydrogen atoms, such as the ones attached to the etheroatoms.}
\end{figure}

In particular, given the total charge density of the solute, the second
term for polarization charge appearing in Eq. (\ref{eq:rhopol_final})
is readily obtained, while for the first one, that we label $\rho^{iter}$:
\begin{equation}
\rho^{iter}=\frac{1}{4\pi}\mathbf{\nabla}\ln\epsilon\left(\rho^{elec}\left(\mathbf{r}\right)\right)\cdot\mathbf{\nabla}\phi^{total}\left(\mathbf{r}\right),\label{eq:rhoiter_def}
\end{equation}
the following procedure can be adopted: 
\begin{enumerate}
\item In order to avoid calculations of polarization effects for a system
with a non converged electronic density, solvent effects are computed
only when the accuracy of the SCF calculation reaches a given threshold
value $\tau^{SCF}$. 
\item At the first iteration for the dielectric, the initial polarization
charge is fixed to be equal to zero. All following calculations at
each subsequent electronic SCF step use the polarization density from
the previous step, namely
\begin{alignat}{1}
\begin{cases}
\rho_{0}^{iter}=0 & \mbox{for the first polarization calculation}\\
\rho_{0}^{iter}=\rho_{old}^{iter} & \mbox{oterwise}
\end{cases} & \mbox{ }\label{eq:rhoiter_zero}
\end{alignat}

\item The total density of the system (solute plus polarization) at iteration
$n$ is computed as
\begin{align}
\rho_{n}^{tot} & \left(\mathbf{r}\right)=\rho^{solute}\left(\mathbf{r}\right)+\rho_{n}^{pol}\left(\mathbf{r}\right)\\
= & \rho^{solute}\left(\mathbf{r}\right)+\rho_{n}^{iter}\left(\mathbf{r}\right)+\frac{1-\epsilon\left(\rho^{elec}\left(\mathbf{r}\right)\right)}{\epsilon\left(\rho^{elec}\left(\mathbf{r}\right)\right)}\rho^{solute}\left(\mathbf{r}\right)\\
= & \frac{1}{\epsilon\left(\rho^{elec}\left(\mathbf{r}\right)\right)}\rho^{solute}\left(\mathbf{r}\right)+\rho_{n}^{iter}\left(\mathbf{r}\right).\label{eq:rhototal}
\end{align}

\item The gradient of the total field is computed in reciprocal space via
one Fast Fourier Transform (FFT) and three inverse FFTs (IFFTs): 
\begin{align}
\rho_{n}^{tot}\left(\mathbf{r}\right) & \overset{FFT}{\rightarrow}\rho_{n}^{tot}\left(\mathbf{g}\right)\label{eq:dfft}\\
\mathbf{\nabla}\phi_{n+1}^{tot}\left(\mathbf{g}\right) & =\frac{4\pi i\mathbf{g}}{g^{2}}\rho_{n}^{tot}\left(\mathbf{g}\right)\label{eq:gradphitotal}\\
\mathbf{\nabla}\phi_{n+1}^{tot}\left(\mathbf{g}\right) & \overset{3\times IFFTs}{\rightarrow}\mathbf{\nabla}\phi_{n+1}^{tot}\left(\mathbf{r}\right).\label{eq:ifft}
\end{align}

\item The iterative part of the polarization charge at the $\left(n+1\right)$-th
step is computed by using Eq. (\ref{eq:rhoiter_def}), namely
\begin{equation}
\rho_{n+1}^{iter}\left(\mathbf{r}\right)=\mathbf{\nabla}\ln\epsilon\left(\rho^{elec}\left(\mathbf{r}\right)\right)\cdot\mathbf{\nabla}\phi_{n+1}^{tot}\left(\mathbf{r}\right).\label{eq:rhoiter_n+1}
\end{equation}

\item A linear mixing of the polarization at the $\left(n+1\right)$-th
and $n$-th steps is performed, in order to stabilize the iterative
procedure:
\begin{gather}
\rho_{n+1}^{iter}\left(\mathbf{r}\right):=\eta\rho_{n+1}^{iter}\left(\mathbf{r}\right)+\left(1-\eta\right)\rho_{n}^{iter}\left(\mathbf{r}\right),\label{eq:mixing}
\end{gather}
with $\eta$ a given mixing parameter (usually $\eta\simeq0.6$),
and the residual $\rho^{res}\left(\mathbf{r}\right)$ is computed
\begin{equation}
\rho^{res}\left(\mathbf{r}\right)=\rho_{n+1}^{iter}\left(\mathbf{r}\right)-\rho_{n}^{iter}\left(\mathbf{r}\right).\label{eq:residual}
\end{equation}

\item The polarization density is converged when 
\begin{eqnarray}
\left\langle \left(\rho^{res}\left(\mathbf{r}\right)\right)^{2}\right\rangle  & < & \tau^{pol},\label{eq:tolerance}
\end{eqnarray}
with $\tau^{pol}$ a given tolerance, and the iterative procedure
is stopped.
\end{enumerate}
Typical results for the polarization density of neutral molecules
in water are reported in Figure (\ref{fig:h2o_c4h10n2}). 

For numerical stability, we find important to compute $\mathbf{\nabla}\ln\epsilon\left(\rho^{elec}\left(\mathbf{r}\right)\right)$
in real space, e.g. by using finite differences. While computing such
a gradient via FFTs would yield a truly variational energy expression,
going through reciprocal space would unavoidably give rise to spurious
polarization charges in {}``forbidden regions''. In particular,
polarization charges close to the nuclei would form, thereby severely
compromising the convergence of our procedure. Several real-space
finite-differences schemes were tested, with the simplest 3-points
central differences algorithm providing converged results for the
majority of tests and simulations. Nonetheless, for microcanonical
molecular dynamics simulations, total energy conservation was found
to strongly depend on the order of the finite-difference scheme adopted,
with only high-order algorithms providing the necessary degree of
accuracy in the interatomic forces (see the discussion in subsection
\ref{sub:Molecular-Dynamics-simulations}). 

Similarly to standard SCF calculations, convergence of iterative calculations
depends crucially on the use of a mixing scheme. Different mixing
schemes can be adopted, such as the Anderson mixing \cite{anderson_jacm_1965},
the modified Broyden mixing \cite{johnson_prb_1988} or the Direct
Inversion in the Iterative Subspace (DIIS) \cite{pulay_cpl_1980}.
In order to reduce the number of numerical parameters involved in
the calculations, in this article only results obtained with a simple
linear mixing are reported. Apart from enhancing convergence, no significant
effects on the final results exist for different mixing parameters
in the range $0.2<\eta<0.6$.

\begin{figure}
\includegraphics[width=0.4\textwidth]{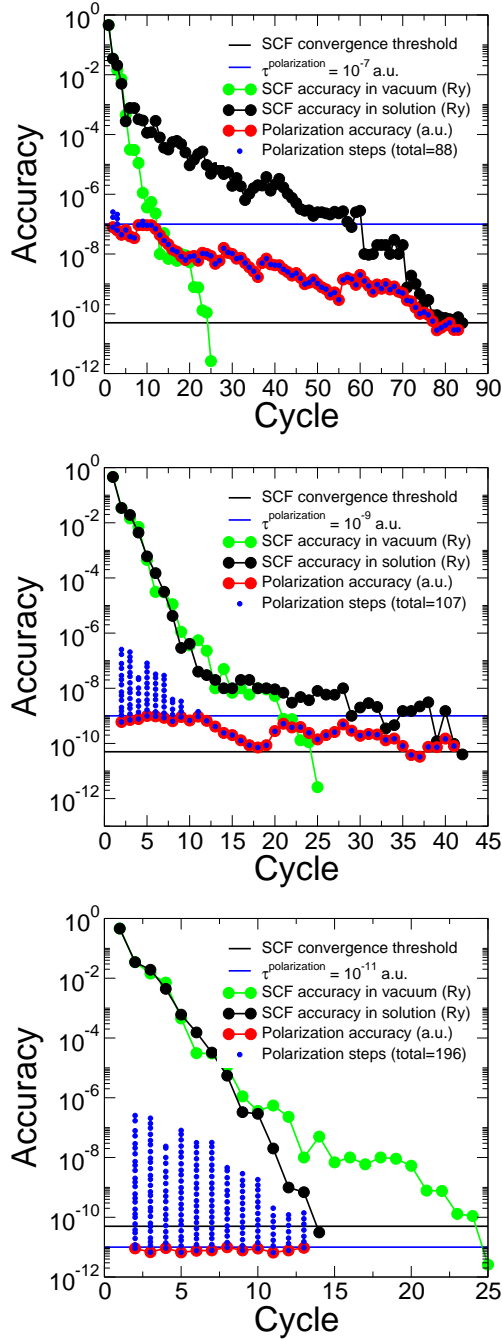}\caption{\label{fig:Effects-of-tolrhopol}Effects of the tolerance on the convergence
of polarization charges $\tau^{pol}$, Eq. (\ref{eq:tolerance}).
Top: when the iterative tolerance is set to a large value, only one
polarization iteration per SCF cycle is needed. The polarization accuracy
increases in unison with the SCF cycles, but convergence of the electronic
density is slowed down. Bottom: for a small value of the iterative
tolerance (i.e. tight convergence in the dielectric cycle), many polarization
iterations are needed for each SCF cycle. Global convergence to the
electronic ground state is not affected by the presence of the solvent,
or actually it is improved with respect to the simulation in vacuum.
Middle: intermediate case.}
\end{figure}

The total number of polarization iterations required is strictly related
to the tolerance $\tau^{pol}$ imposed on the residuals. A large value
of $\tau^{pol}$ in Eq. (\ref{eq:tolerance}) would allow the polarization
procedure to stop in a few iterations. Nonetheless, poorly converged
polarization screening does prevent the electrons from reaching their
ground state, requiring a much higher number of SCF steps to converge
(see top and middle panels of Figure (\ref{fig:Effects-of-tolrhopol})).

Since a single SCF step is much more computationally expensive than
a polarization iteration, which only involves four FFTs, it has been
found preferable to impose a tight convergence tolerance on the polarization
charge (bottom panel of Figure (\ref{fig:Effects-of-tolrhopol})).
More advanced schemes that rely on a tolerance that depends on the
convergence of the electronic density, have been tested. Moreover,
a scheme could be envisaged where the polarization density and the
electronic density are optimized with the same common iterative algorithm,
i.e., where the polarization density is mixed once per SCF cycle and
with the same procedure used for the electronic density. Nonetheless,
for sake of simplicity, only results obtained with a fixed tolerance
on the polarization charge are presented.

The behavior of the SCF procedure is also strictly dependent on the
choice of the threshold value for the onset of the polarization calculation,
$\tau^{SCF}$. Waiting for the electrons to be converged before starting
the polarization calculation could seem an efficient way of proceeding.
Nonetheless, switching on the polarization fields strongly affects
the SCF procedure, bringing the electrons as far from their ground
state as at first SCF steps. For this reason, larger values of $\tau^{SCF}$,
of the order of 0.1-0.5 Ry, were found to give smoother SCF convergence,
with no effects on the final results.

Overall, the whole procedure benefits significantly from the advantageous
scaling of FFT techniques. Contrary to multigrid solvers, FFTs are
easily available in fast, well established, parallel libraries and
form already a key ingredient in most plane-wave electronic-structure
codes. Moreover, the method relies on fully periodic boundary conditions,
so it can be straightforwardly applied to periodic systems in arbitrary
unit-cell geometries without requiring a cubic cell and imposing arbitrary
Dirichlet boundary conditions on the potential, as in the multigrid
cases. In polar solvents, periodicity does not affect final results,
since the solvent screens very effectively the charge density of the
solute. Moreover, periodic image correction schemes, such as the Makov-Payne
correction \cite{makov_prb_1995} or countercharge methods \cite{dabo_prb_2008,li_prb_2011},
can easily be adopted to include the polarization charge density,
by treating it at the same level as the molecular charge distribution.

\section{\label{sec:Extra-terms}Additional solvation terms}

The model described in the previous sections focuses only on electrostatic
contributions to solvation \cite{tomasi_chemrev_2005} 
\begin{equation}
\Delta G^{el}=G^{el}-G^{0},\label{eq:dg_electrostatic}
\end{equation}
where $G^{0}\equiv\left(E^{tot}\right)_{vacuum}$ is the ab-initio
energy of the isolated solute in vacuum and $G^{el}$ is the analogous
quantity computed in solution, i.e. from Eq. (\ref{eq:kohn-sham_energy}).
Such a contribution is always negative, i.e., allowing the medium
to polarize always stabilizes the solute and lowers its total energy.
Of course, other effects than electrostatic are present in any solvation
process and they play a crucial role in balancing the overall solvation
energies that can be negative or positive. According to the formal
definitions of Ben-Naim \cite{ben-naim1987,ben-naim1992}, PCM introduced
other non-electrostatic terms in the solute Hamiltonian, still assuming
a continuum approach in the description of the medium. The main terms
can be divided (see Eq. (79) of Ref. \cite{tomasi_chemrev_2005})
into: 
\begin{equation}
\Delta G^{sol}=\Delta G^{el}+G^{cav}+G^{rep}+G^{dis}+\Delta G^{tm}+P\Delta V.\label{eq:dgsol_full}
\end{equation}
In the above formula, the cavitation energy $G^{cav}$ corresponds
to the energy necessary to build into the solvent the cavity containing
the solute. The repulsion $G^{rep}$ and dispersion $G^{dis}$ terms
are the continuum equivalent of the nonbonded short-range interactions
generated by the Pauli exclusion principle and by van der Waals interactions.
The thermal motion contribution $G^{tm}$ arises from the change in
the vibrational and rotational properties of the solvated system with
respect to the isolated one, and the pressure term $P\Delta V$ takes
into account the change in volume of the solvated system. Different
models exist for each of the non-electrostatic terms, a comprehensive
review being given in Ref. \cite{tomasi_chemrev_2005}. In standard
PCM calculations only the first three non electrostatic terms are
explicitly accounted for. Of these, cavitation and repulsion energies
are positive by definition, while the dispersion contribution is always
negative. Overall, these terms tend to cancel each other and their
effect on molecular properties and chemical reactions has been generally
regarded as less important than the electrostatic term.

The model by Fattebert and Gygi was extended by Scherlis et al. \cite{scherlis_jcp_2006}
to include a cavitation term, simply expressed as the product of the
experimental surface tension of the solvent $\gamma$ times the surface
$S$ of the solute cavity, namely
\begin{equation}
G^{cav}=\gamma S.\label{eq:g_cavitation}
\end{equation}
The surface above is the {}``quantum surface'' introduced in \cite{cococcioni_prl_2005}
and defined, via the {}``quantum volume'' \cite{cococcioni_prl_2005},
by finite difference between two isosurfaces of the electronic density
as

\begin{equation}
S=\int d\mathbf{r}\left\{ \vartheta_{\rho_{0}-\frac{\Delta}{2}}\left(\rho^{elec}\left(\mathbf{r}\right)\right)-\vartheta_{\rho_{0}+\frac{\Delta}{2}}\left(\rho^{elec}\left(\mathbf{r}\right)\right)\right\} \times\frac{\left|\mathbf{\nabla}\rho^{elec}\left(\mathbf{r}\right)\right|}{\Delta}.\label{eq:surface_scherlis}
\end{equation}
From the functional derivative of the cavitation energy (Eq. (\ref{eq:g_cavitation}))
with respect to the charge density, an extra potential term of the
form 
\begin{align}
\frac{\delta G^{cav}\left[\rho^{elec}\right]}{\delta\rho^{elec}} & =\frac{\gamma}{\Delta}\left\{ \vartheta_{\rho_{0}-\frac{\Delta}{2}}\left(\rho^{elec}\left(\mathbf{r}\right)\right)-\vartheta_{\rho_{0}+\frac{\Delta}{2}}\left(\rho^{elec}\left(\mathbf{r}\right)\right)\right\} \nonumber \\
\times & \left[\sum_{i}\sum_{j}\frac{\partial_{i}\rho^{elec}\left(\mathbf{r}\right)\partial_{j}\rho^{elec}\left(\mathbf{r}\right)\partial_{i}\partial_{j}\rho^{elec}\left(\mathbf{r}\right)}{\left|\nabla\rho^{elec}\left(\mathbf{r}\right)\right|^{3}}-\sum_{i}\frac{\partial_{i}^{2}\rho^{elec}\left(\mathbf{r}\right)}{\left|\nabla\rho^{elec}\left(\mathbf{r}\right)\right|}\right]\label{eq:cavitation_potential}
\end{align}
can be straightforwardly added to the Kohn-Sham Hamiltonian. The $\vartheta_{\rho_{0}}$
function entering in the above equations was originally designed to
be a step function, switching from zero to one at the threshold $\rho_{0}$,
namely,
\begin{equation}
\vartheta_{\rho_{0}}\left(\rho^{elec}\left(\mathbf{r}\right)\right)=\vartheta\left(\rho^{elec}\left(\mathbf{r}\right)-\rho_{0}\right)=\begin{cases}
0 & \rho^{elec}\left(\mathbf{r}\right)-\rho_{0}>0\\
1 & \rho^{elec}\left(\mathbf{r}\right)-\rho_{0}<0
\end{cases}.\label{eq:theta_cococcioni}
\end{equation}
To improve the numerical stability of the algorithm, a smoothed step
function of the same form, i.e. $\vartheta\left(\rho^{elec}\left(\mathbf{r}\right)-\rho_{0}\right)$,
was also suggested. The finite difference parameter $\Delta$ was
then used in Eq. (\ref{eq:surface_scherlis}) to determine two adjacent
isosurfaces separated by a constant separation in terms of electronic
density. 

Subsequently, in order to use a definition consistent with the dielectric
function used in the electrostatic solvation calculation, the function
\begin{equation}
\vartheta_{\rho_{0},\beta}\left(\rho^{elec}\left(\mathbf{r}\right)\right)=\frac{1}{2}\left[\frac{\left(\rho^{elec}\left(\mathbf{r}\right)/\rho_{0}\right)^{2\beta}-1}{\left(\rho^{elec}\left(\mathbf{r}\right)/\rho_{0}\right)^{2\beta}+1}+1\right]\label{eq:theta_scherlis}
\end{equation}
was adopted. We note that, while very similar in spirit to the original
formulation by Cococcioni et. al. \cite{cococcioni_prl_2005}, the
use of Eq. (\ref{eq:theta_scherlis}) together with Eq. (\ref{eq:surface_scherlis})
is not formally exact, since the parameter $\rho_{0}$ entering in
Eq. (\ref{eq:theta_scherlis}) is not in a linear relationship with
the argument of the switching function. Thus, applying the finite
difference procedure to this arbitrary parameter as in Eq. (\ref{eq:surface_scherlis})
does not correspond to calculating the spatial distance between the
two isosurfaces. For this reason, surfaces computed with Eq. (\ref{eq:surface_scherlis})
result in an overestimate of approximately a factor of 1.2 of the
correct quantum surface. Such an overestimate can be avoided by applying
the finite difference procedure directly to the argument of the switching
function, namely, 
\begin{equation}
S=\int d\mathbf{r}\left\{ \vartheta_{\left\{ \alpha\right\} }\left(\rho^{elec}\left(\mathbf{r}\right)-\frac{\Delta}{2}\right)-\vartheta_{\left\{ \alpha\right\} }\left(\rho^{elec}\left(\mathbf{r}\right)+\frac{\Delta}{2}\right)\right\} \times\frac{\left|\mathbf{\nabla}\rho^{elec}\left(\mathbf{r}\right)\right|}{\Delta}.\label{eq:surface_correct}
\end{equation}
where now the $\vartheta_{\left\{ \alpha\right\} }\left(\rho^{elec}\left(\mathbf{r}\right)\right)$
function is a given switching function that depends on an arbitrary
set of parameters $\left\{ \alpha\right\} $ and that goes from zero
to one at a fixed threshold larger than $\Delta/2$. We notice here
that, while the above expression is valid for any kind of switching
function, for a $\vartheta$ function defined according to the original
formulation of Cococcioni et al. {[}Eq. (\ref{eq:theta_cococcioni}){]},
Eqs. (\ref{eq:surface_scherlis}) and (\ref{eq:surface_correct})
are equivalent. For the sake of consistency with the electrostatic
solvation term the following definition is adopted 
\begin{equation}
\vartheta_{\rho_{min},\rho_{max}}\left(\rho^{elec}\left(\mathbf{r}\right)\right)=\frac{\epsilon_{0}-\epsilon_{\epsilon_{0},\rho_{min},\rho_{max}}\left(\rho^{elec}\left(\mathbf{r}\right)\right)}{\epsilon_{0}-1},\label{eq:theta_correct}
\end{equation}
where $\epsilon_{\epsilon_{0},\rho_{min},\rho_{max}}\left(\rho^{elec}\left(\mathbf{r}\right)\right)$
is now given by Eq. (\ref{eq:eps_new}). 

In a similar way, the method of Cococcioni et al. to treat systems
under pressure can be immediately extended to the calculation of the
$P\Delta V$ term that appears in Eq. (\ref{eq:dgsol_full}). Explicitly,
by adopting the same switching function used to define the solute
surface Eq. (\ref{eq:theta_correct}), the solute volume can be expressed
as 
\begin{equation}
V=\int d\mathbf{r}\vartheta\left(\rho^{elec}\left(\mathbf{r}\right)\right),\label{eq:volume}
\end{equation}
that gives rise to an extra potential term in the Kohn-Sham Hamiltonian
of the form 
\begin{equation}
\frac{\delta PV\left[\rho^{elec}\right]}{\delta\rho^{elec}}=P\frac{\delta\vartheta\left(\rho^{elec}\right)}{\delta\rho^{elec}}.\label{eq:volume_potential}
\end{equation}
By including the $PV$ term in the total energies of both the system
in vacuum and in solution, the $P\Delta V$ contribution is automatically
included in the solvation free energy computed via Eq. (\ref{eq:dg_electrostatic}).
Such a contribution can be important for systems under pressure while,
for systems at standard pressures, it can be safely neglected. 

As for the remaining contributions to the solvation free energy, we
have decided to treat them in a simplified way, their explicit modeling
being the subject of future developments. In particular, similarly
to other models of solvation, the thermal motion contribution has
been neglected, while we express the sum of dispersion and repulsion
free energies as a term linearly proportional to the quantum surface
and the quantum volume of the molecular cavity, namely
\begin{equation}
G^{rep}+G^{dis}=\alpha S+\beta V,\label{eq:g_rep_g_dis}
\end{equation}
where the two factors $\alpha$ and $\beta$ are solvent-specific
tunable parameters that can be fitted, together with the other parameters
in the model, e.g. to reproduce total solvation energies. This approach
is coherent, for example, with the definition of repulsion energy
used in PCM \cite{amovilli_jpcb_1997}, where such a term is proportional
to the solute electronic density that lies outside the molecular cavity:
in a model where the molecular cavity is defined on an isosurface
of the density, the amount of electronic charge outside the cavity
will be proportional to the surface of the cavity. Moreover, such
an approach is similar in spirit, but with less parameters, to the
one adopted by Cramer and Truhlar in their Generalized Born (GB) method
of solvation and, more recently, in their SMx models (see Ref. \cite{cramer_accchemres_2008}
and references therein). In these classes of methods, all of the non-electrostatic
terms of solvation are described as the sum of atomic terms, proportional
to the solvent-accessible surface area of each atom of the solute. 

The final expression for the solvation free energy of our simplified
model is thus given by
\begin{equation}
\Delta G^{sol}=\Delta G^{el}\left(\epsilon_{0},\rho_{min},\rho_{max}\right)+\left(\alpha+\gamma\right)S+\beta V,\label{eq:dgsol_simple}
\end{equation}
where the main parameters involved in the electrostatic contribution
are explicitly reported.

\section{\label{sec:Results}Results}

\subsection{\label{sub:Numerical-Details}Numerical Details}

\begin{figure}
\includegraphics[width=0.8\textwidth]{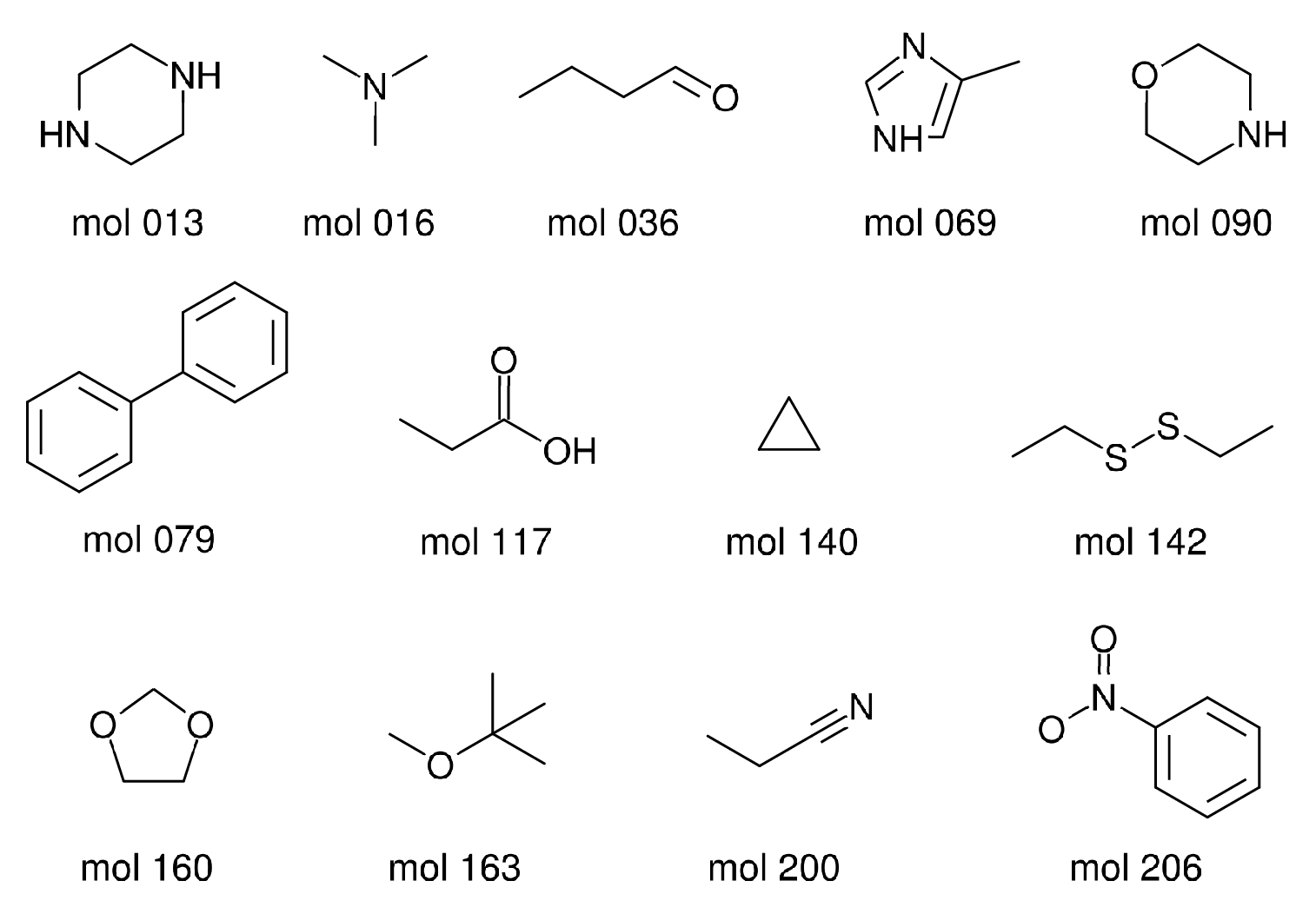}\caption{\label{fig:Fitting-set}Fitting set: chemical structures and labels
of the thirteen small, neutral, organic molecules used to fit the
solvation model and to test the accuracy of the model with respect
to numerical parameters. Names, solvation energies and computed results
for the molecules in the set are reported in Tables in the Supplementary
Material \cite{suppmat}.}
\end{figure}

A large set of experimental solvation free energies for 240 small
neutral organic molecules in water was used to test the proposed computational
methodology. The molecules of the set have been collected in Ref.
\cite{shivakumar_jctc_2010} to test the accuracy of classical molecular
dynamics free energy perturbation calculations. Based on hierarchical
clustering, a smaller representative set of 13 molecules (reported
in Figure \ref{fig:Fitting-set}) spanning the main functional groups
of organic molecules is also provided in Ref. \cite{shivakumar_jctc_2010}
and is used here to test the convergence of numerical parameters and
to fit tunable parameters.

PCM calculations were performed at the Hartree-Fock (HF) and density
functional theory (DFT) level, using both the G03 \cite{g03} and
G09 \cite{g09} versions of the GAUSSIAN code. For the sake of brevity,
in the remaining discussions the labels g03/g09 will be used to refer
to the two different flavors of PCM \cite{mennucci_jpcb_1997,scalmani_jcp_2010}
implemented as defaults in the G03/G09 versions of the GAUSSIAN code.
Both standard PBE and hybrid B3LYP exchange-correlation functionals
were tested. The dependence of the results on the chosen theoretical
approximations was found to be minor and generally negligible compared
to the overall agreement with experimental results. Thus, only results
obtained with the PBE functional will be reported. Geometries were
optimized both in vacuum and in solution with a double zeta 6-31g(d)
basis set. Starting from the relaxed geometries, total energies were
computed with a triple zeta 6-311+g(d,p) basis set. Due to the lack
of such basis sets for the iodine atom, calculations with GAUSSIAN
were not performed on the eight molecules of the set containing this
atomic species. For the other molecules, calculations were performed
using PCM to compute both electrostatic and cavitation energies. For
the latter contribution and in order to be coherent with the original
definition of the cavitation term, calculations were performed with
a cavity built using unscaled Bondi's atomic radii \cite{bondi_jpc_1964}.
For all other calculations the default cavities were used. In particular,
IEF-PCM as implemented in G03 adopts a cavity defined according to
the solvent excluded surface \cite{connolly_jac_1983} as approximated
by the GePol algorithm \cite{pascualahuir_jcc_1994}, and based on
spheres of United Atom radii \cite{barone_jcp_1997}. The version
of IEF-PCM implemented in G09, instead, relies on a simpler cavity
\cite{scalmani_jcp_2010}, built as the combination of atom centered
van der Waals spheres of Universal Force-Field radii \cite{rappe_jacs_1992},
and scaled by a factor of 1.1. 

Our continuum solvation model has been implemented in the public-domain
PWSCF parallel code included in the Quantum-ESPRESSO package \cite{QUANTUM-espresso},
based on DFT, periodic-boundary conditions, plane-wave basis sets
and pseudopotentials (PP) to represent the ion-electron interactions.
All the calculations reported in this work were performed at the Gamma
point of the Brillouin zone (as is appropriate for molecules), using
the PBE exchange-correlation functional and Vanderbilt ultrasoft pseudopotentials
(USPP), as contained in the PSlibrary of A. Dal Corso \cite{pslibrary}.
For bromine and iodine we have used the USPP available online \cite{pseudo-QE}.
Kohn-Sham wavefunctions and charge densities were expanded in plane
waves up to kinetic energy cutoffs of 30 and 300 Ry, respectively.
For molecules containing fluorine, these cutoffs were further raised
to 45 and 450 Ry. Simulation cells were chosen to be at least 15.0
a.u. larger than the maximum size of the molecule in vacuum, and not
smaller than 20.0 a.u.. Spurious periodic-image effects were taken
into account, for all the molecules, using the Makov-Payne corrective
scheme \cite{makov_prb_1995}. 

In order to ensure a consistent description of solvent effects across
our large set of simulations used to fit the tunable parameters of
the method, a larger basis set was adopted, corresponding to wave-function
and density cutoffs of 40 and 400 Ry, respectively. Moreover, since
the effect of the tunable parameters on the final geometries of the
benchmark molecules was found to be of secondary importance (less
than 0.1 kcal/mol), the fitting simulations were performed with no
geometry relaxation and starting from molecular geometries optimized
using a reasonable set of parameters ($\rho^{max}=0.008$ a.u., $\rho^{min}=0.00015$
a.u., $\gamma=72$ dyn/cm, $\alpha=0$ dyn/cm, $\beta=0$ GPa).

The accuracy of the method and, in particular, of the forces calculation
were tested by means of Born-Oppeneimer ab-initio molecular dynamics
simulations of a prototypical system in vacuum and in solution. Equilibration
of the system was first performed with a timestep of 60 a.u. ($\approx1.45$
fs) in the NVT ensemble at 300 K, as imposed by a Berendsen thermostat.
Following equilibration at constant temperature, simulations were
performed in the microcanonical ensemble. In order to ensure proper
energy conservation for the simulation in vacuum, some key simulations
parameters had to be tightened. In particular, the wavefunction and
density cutoffs were raised to 40 Ry and 600 Ry, respectively. Similarly,
the timestep was decreased to 20 a.u. ($\approx0.48$ fs) and the
threshold controlling the convergence of the electronic SCF procedure
was imposed to be $10^{-11}$ Ry. The same set of parameters was then
used for the simulations in the presence of the solvent.

In order to study the influence of the numerical parameters of the
method on the final results, unrelaxed electrostatic solvation energies
were calculated as
\begin{equation}
\Delta G^{el,0}=G^{el,0}-G^{0},\label{eq:dg_electrostatic_unrelaxed}
\end{equation}
where, contrary to Eq.(\ref{eq:dg_electrostatic}), the electrostatic
energy of the system in solution is computed using the geometry optimized
in vacuum ($G^{el,0}$). This was done with the aim of removing a
further source of errors from the analysis of the numerical sensitivity
of the method.

\subsection{Parametrization}

As summarized in Eq. (\ref{eq:dgsol_simple}), our approach involves
six main parameters. Of these, the bulk dielectric constant $\epsilon_{0}$
and the surface tension $\gamma$ represent physical quantities, specific
of the solvent, that can be computed from first principles or, more
often, extracted from experimental results. The remaining four parameters,
i.e. the two thresholds entering into the definition of the dielectric
function and the $\alpha$ and $\beta$ constants of Eq. (\ref{eq:g_rep_g_dis}),
are tunable quantities that affect the computed solvation energies.
As such, these parameters have to be fitted on reference calculations
or experimental data, as reported in the following subsection (\ref{sub:Fitting-of-tunable}).
It is important to underline here that the comparisons with PCM are
not intended to represent a direct assessment of the accuracy of PCM
or of other continuum models in the literature, but rather they serve
two other purposes. First, they are intended to show the intrinsic
flexibility of the SCCS formulation, which can reproduce well results
of other approaches exploiting a much reduced number of parameters.
Second, using well established continuum models simplifies the fitting
procedure in comparison to relying solely on experimental data. For
the above reasons and due to their wide spread use and assessment
in the literature, comparisons and fittings are only made with the
two main versions of IEF-PCM, as implemented in the Gaussian simulation
packages \cite{cossi_jcp_2002,scalmani_jcp_2010}. We are aware that
other continuum models exist in the literature \cite{klamt_jchemsoc_1993,wiberg_jpc_1995,zhan_jcp_1998,tomasi_chemrev_2005,cramer_accchemres_2008}
and some of these \cite{wiberg_jpc_1995,zhan_jcp_1998} could also
be fitted within the SCCS formulation; nonetheless such extensive
comparison is beyond the purpose of this article.

In addition to the parameters reported in Eq. (\ref{eq:dgsol_simple}),
the proposed methodologies rely on a certain number of computational
parameters. These are purely numerical quantities, that influence
the speed and stability of the calculations but, apart from pathological
cases, should not affect the calculated results. These parameters
include the standard parameters of plane-wave periodic boundary calculations
(wavefunction and density cutoffs, cell size, pseudopotentials) together
with some solvent specific quantities, such as the tolerances of the
iterative approach $\tau^{SCF}$ and $\tau^{pol}$, the parameter
$\eta$ entering into the polarization density mixing approach, the
finite difference parameter $\Delta$ used in the calculation of the
molecular surface, and the radii of the fictitious ionic densities
used in the Ewald evaluation of electrostatic contributions. While,
in theory, these parameters can be freely adjusted to make the method
more efficient and more stable, a careful analysis of their effects
on the final results is mandatory and is reported in Appendix A (\ref{sub:Effect-of-numerical}).

\subsubsection{\label{sub:Fitting-of-tunable}Fitting of tunable parameters}

In particular, since the proposed model mostly focuses on the electrostatic
contribution to solvation free energies, it is difficult and even
not physical to fit tunable parameters of the model to reproduce total
solvation free energies of molecules. Indeed, such a fit would intrinsically
rely on cancellation of errors and would be highly dependent on the
solutes and solvents considered. Furthermore, it is mostly the electrostatic
part of solvation that plays a role in determining the molecular properties
of a solute in solution, such as solvent effects on molecular spectra
(IR, UV or NMR) or on reaction rates. An ideal fitting would thus
require parametrizing the model on these molecular properties. Since
already existing continuum models of solvation have generally achieved
a very good description of molecular properties in solution, and the
IEF-PCM predictions for electrostatic solvation energies have been
successfully used by several methods \cite{cramer_accchemres_2008,curutchet_jcomputchem_2001,soteras_jmolstrtheo_2005,klamt_accchemres_2009}
as the basis onto which one could parametrize more complete models
of solvation, we decided to use the electrostatic part of the results
obtained with the PCM to parametrize the electrostatic part of our
model, while the non-electrostatic terms have been tuned in a second
stage in order to improve the fit with respect to experimental total
solvation free energies. 

The two main parameters of the electrostatic part of SCCS to be fitted
are the two density thresholds $\rho^{min}$ and $\rho^{max}$ that
enter into the definition of the dielectric constant {[}Eq. (\ref{eq:eps_new}).
In particular, $\rho^{max}$ defines the onset of the dielectric and,
for the flatness conditions (see Section \ref{sec:Choice-of-dielectric})
to be fulfilled, should correspond to a density isosurface as far
from the nuclei as possible. Thus we consider only values lower than
0.005 a.u. for $\rho^{max}$. As for $\rho^{min}$, since the electronic
density in plane-wave codes is computed via FFTs and can present small
oscillations also in regions of space far from the molecule, it is
important to define this threshold such that it is not influenced
by numerical noise in the electronic density. For this reason only
values of $\rho^{min}$ higher than 0.00005 a.u. have been considered
in the fit. 

Mean absolute errors (MAE) relative to PCM electrostatic solvation
energies as a function of $\rho^{min}$ and $\rho^{max}$ are reported
in Figures (\ref{fig:Fit-of-G09}) and (\ref{fig:Fit-of-G03}), for
G09 and G03 respectively, and in the Supplementary Material \cite{suppmat}.
From both graphs, it appears that there is no unique optimal fit for
the two parameters $\rho^{min}$ and $\rho^{max}$. In fact, for each
value of $\rho^{min}$ there exists a value of $\rho^{max}$ that
provides the optimal fit with respect to the reference. This observation
offers some additional freedom in the choice of the two parameters. 

The agreement with the G09 results is excellent with a mean absolute
error (MAE) lower than 0.4 kcal/mol on the 13 benchmark molecules
of the training set (see Figure (\ref{fig:Fit-of-G09}a)). Out of
the possible choices of the two parameters, the values of $\rho^{min}=0.0001$
a.u. and $\rho^{max}=0.0015$ a.u. were selected (fitg09), giving
a minimal MAE of 0.36 kcal/mol for the 13 molecules of the training
set. The MAE is found to further decrease to 0.27 kcal/mol for the
full set of 240 molecules reflecting the remarkable transferability
of the fit performed on only 13 benchmark molecules. This results
is made more remarkable by the fact that our model relies on only
two parameters for the definition of the cavity hosting the solute,
at variance with PCM that involves in its simplest implementation
a different empirical parameter for each atomic species.

\begin{figure}
\includegraphics[width=0.8\textwidth]{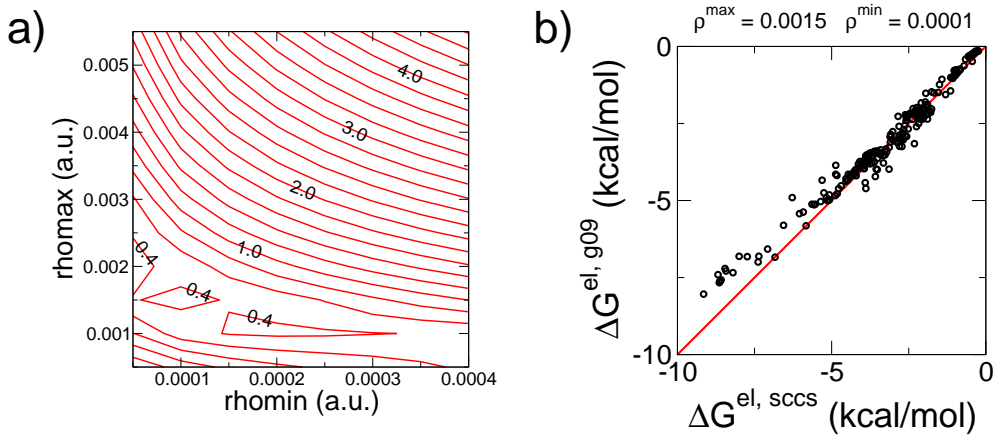}\caption{\label{fig:Fit-of-G09}a) SCCS mean absolute error with respect to
PCM-G09 electrostatic solvation energies for the 13 molecules of the
training set. b) Comparison of SCCS electrostatic predictions with
PCM-G09 predictions for the 240 molecules of the full set. }
\end{figure}

Parametrizing the method on the electrostatic solvation energies obtained
with the G03 version of the PCM model resulted in a less close agreement,
with MAE of the order of 1.0 kcal/mol (see Figure (\ref{fig:Fit-of-G03}a));
again, also in this case, results improve when the full set of 240
neutral molecules is considered. In particular, using $\rho^{min}=0.0001$
a.u. and $\rho^{max}=0.0050$ a.u. (fitg03), an average error of 0.95
kcal/mol is obtained (see Figure (\ref{fig:Fit-of-G03}b)). By comparing
Figures (\ref{fig:Fit-of-G09}a) and (\ref{fig:Fit-of-G03}a), it
is important to notice that, for each value of $\rho^{min}$, the
value of $\rho^{max}$ that gives the best fit of the PCM results
in the G03 case is always higher than the corresponding G09 fit. This
corresponds to a cavity that is closer to the solute, and is a behavior
to be expected. The reason for such a result lies in the different
definition of the molecular cavity in the two version of the model,
the G09 version relying, in the default implementation, on a larger
solute cavity \cite{scalmani_jcp_2010}. Similarly, the reason for
the less good agreement between SCCS and PCM G03 results is most likely
due to the more elaborate definition of the molecular cavity in PCM
G03, where an additional set of spheres was added to the cavity to
smooth out intersecting regions. 

\begin{figure}
\includegraphics[width=0.8\textwidth]{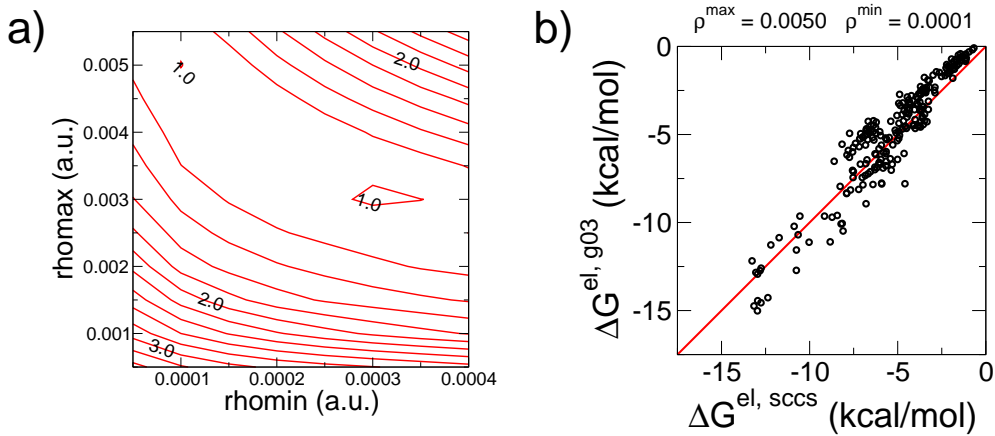}\caption{\label{fig:Fit-of-G03}a) SCCS mean absolute error with respect to
PCM-G03 electrostatic solvation energies for the 13 molecules of the
training set. b) Comparison of SCCS electrostatic predictions with
PCM-G03 predictions for the 240 molecules of the full set. }
\end{figure}

\begin{figure}
\includegraphics[width=0.8\textwidth]{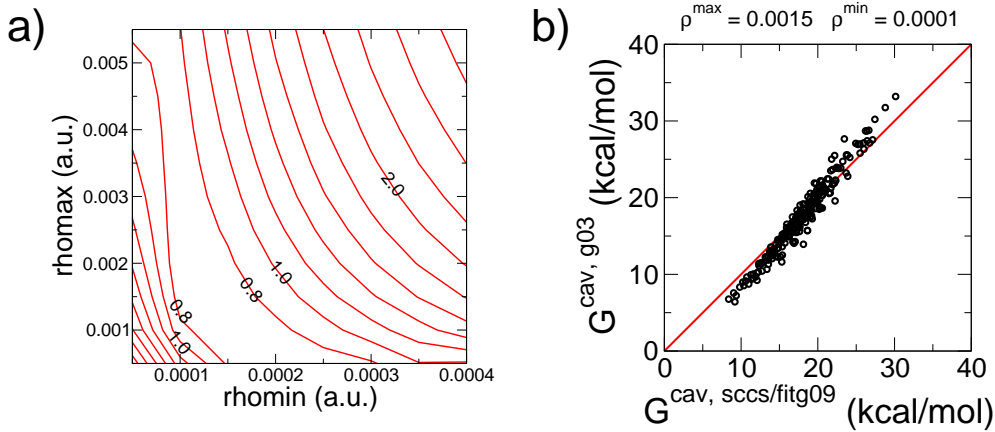}

\caption{\label{fig:Fit-of-cavitation}a) SCCS mean absolute error with respect
to cavitation energies computed by PCM-G03 for the 13 molecules of
the training set. b) Comparison of SCCS cavitation predictions with
PCM-G03 predictions for the 240 molecules of the full set, using the
cavity parameters optimized to reproduce the electrostatic solvation
energy of PCM-G09. The value of $\gamma$ is set to be equal to 72
dyn/cm. }

\end{figure}

As a further test of the flexibility of the proposed SCCS model, a
comparison of the cavitation energies computed with Eq. (\ref{eq:g_cavitation})
versus that of the Claverie-Pierotti method implemented in PCM was
performed and reported in Figure (\ref{fig:Fit-of-cavitation}). Also
in this case, even though the two methods are based on different physical
assumptions, results show a remarkable agreement. Errors of the order
of 0.8 kcal/mol were found on the computed cavitation energies, despite
the fact that they are typically four times larger than solvation
energies. Calculations of $G^{cav}$ on the whole set of molecules
were performed using the experimental value of $\gamma=72$ dyn/cm
for liquid water at room temperature and the fitg09 set of parameters,
since these set lie close to the region of best match with the PCM
cavitation energies (see Figure (\ref{fig:Fit-of-cavitation})). 

\begin{figure}
\includegraphics[width=0.8\textwidth]{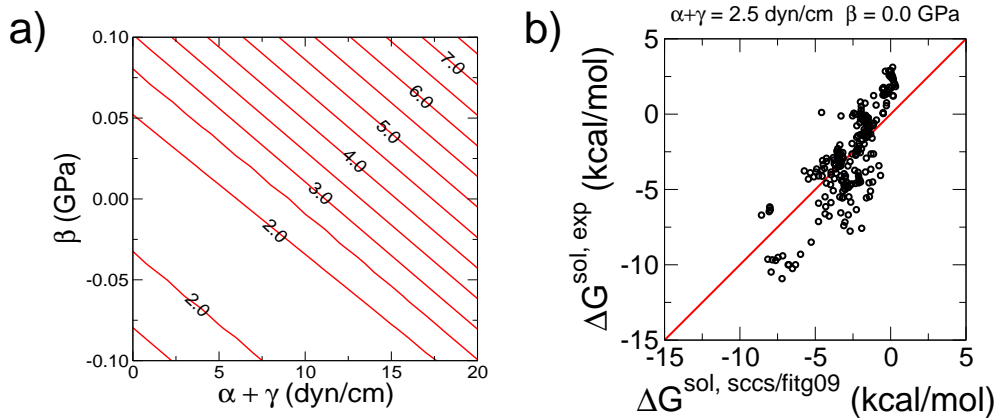}

\caption{\label{fig:Nonelectrostatic-Fit-G09}a) SCCS mean absolute error with
respect to experiments for the 13 molecules of the training set. b)
Comparison of SCCS solvation energies with experimental solvation
energies for the 240 molecules of the full set. In both graphs, the
electrostatic parameters of the SCCS are fitted on G09 ($\rho^{min}=0.0001$
a.u., $\rho^{max}=0.0015$ a.u.). Results in b) are obtained with
the fitg09 set of parameters (see Table (\ref{tab:Mean-absolute-error})).}

\end{figure}

\begin{figure}
\includegraphics[width=0.8\textwidth]{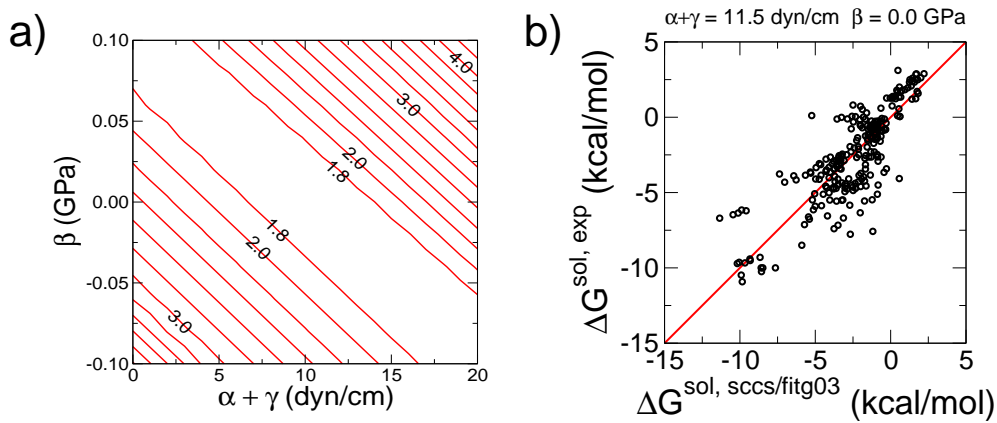}\caption{\label{fig:Nonelectrostatic-Fit-G03}a) SCCS mean absolute error with
respect to experiments for the 13 molecules of the training set. b)
Comparison of SCCS solvation energies with experimental solvation
energies for the 240 molecules of the full set. In both graphs, the
electrostatic parameters of the SCCS are fitted on G03 ($\rho^{min}=0.0001$
a.u., $\rho^{max}=0.0050$ a.u.). Results in b) are obtained with
the fitg03 set of parameters (see Table (\ref{tab:Mean-absolute-error})).}

\end{figure}

Eventually, while keeping the two parameters of the cavity fixed to
either the fitg09 or fitg03 values, and assuming a surface tension
$\gamma=72$ dyn/cm for water at room temperature, the parameters
$\alpha$ and $\beta$ entering into the definition of the non-electrostatic
terms, Eq. (\ref{eq:g_rep_g_dis}), have been optimized to best reproduce
the experimental results of solvation free energies (see Figures (\ref{fig:Nonelectrostatic-Fit-G09})
and (\ref{fig:Nonelectrostatic-Fit-G03})). 

\begin{figure}
\includegraphics[width=0.5\textwidth]{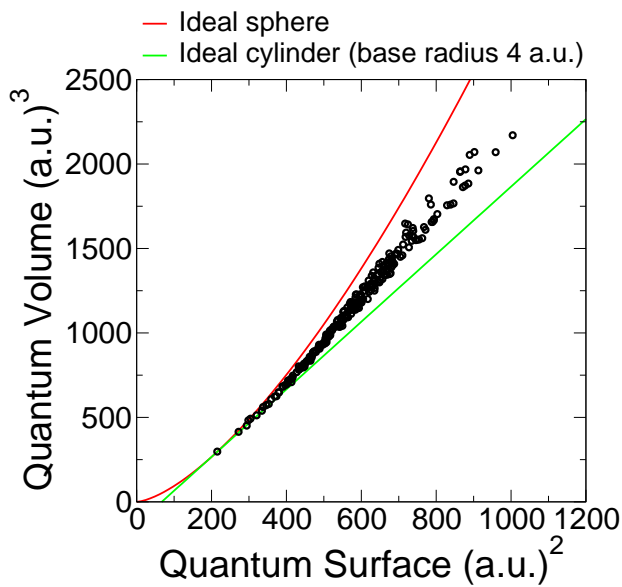}\caption{\label{fig:cav_vs_pres}Relation between the quantum surfaces and
the quantum volumes of the 240 molecules of the full set, computed
using SCCS with the fitg03 set of parameters. For comparison, the
volume vs surface trends are reported for the cases of a spherical
system (in red) and of a cylindrical system terminated by hemispheres
and of fixed base radius equal to 4 a.u. (in green). }

\end{figure}

In both cases, an almost perfectly linear behavior was found for the
error as a function of the two parameters. This is probably due to
the small variation in the size of the molecules considered in the
training set, which makes the volume and the surface terms to be linearly
related (see Figure (\ref{fig:cav_vs_pres})). When looking at the
behavior of the quantum volume with respect to the quantum surface
of the solute for the 240 molecules of the full set, we found a trend
in between the one expected for an ideally spherical system, for which
\begin{equation}
V\propto S^{3/2},
\end{equation}
and a linear system, composed by a cylinder of length $l$, terminated
by hemispheres, and of radius equal to an average atomic size ($r=4$
a.u.), for which
\begin{equation}
V\propto S.
\end{equation}

 As a result of this almost linear behavior that reflects the corrugation
of the solvation shell, we are again left with the freedom of choosing
one of the two parameters, while performing an optimal fit on the
other. In an effort to simplify the method and conform to other solvation
models in the literature, we decided to choose the sum of the surface
parameters $\alpha$ and $\gamma$ such that the volume contribution
vanishes, i.e. in such a manner that the corresponding optimal value
of $\beta$ is equal to zero. This choice corresponds to a value of
$\alpha+\gamma=2.5$ dyn/cm for the fitg09 set of parameters, or a
value of $\alpha+\gamma=11.5$ dyn/cm in the fitg03 case. Total solvation
energies for the 13-molecules training set with these choices of parameters
resulted in a MAE with respect to experimental results of about 1.6
kcal/mol (see Tables in the Supplementary Material \cite{suppmat}).
When considering the whole set of neutral molecules, it is clear that
the fitg03 set of parameters presents the best trend with respect
to experimental data. This results in a better agreement with experiment
for the fitg03 set compared to the fitg09, with MAEs of 1.31 kcal/mol
and 1.53 kcal/mol respectively.

Other fits were performed and tested on the full set of molecules,
to check the consistency of the results along different combinations
of parameters (see Table (\ref{tab:Mean-absolute-error})). It is
instructive to compare two choices of parameters that best reproduce
the electrostatic energies computed with G03, namely the fitg03 and
fitg03' sets in Table (\ref{tab:Mean-absolute-error}). In this case,
it appears that MAEs of comparable magnitude on the 13-molecule training
set (see Figure (\ref{fig:Fit-of-G03})) correspond to similar errors
on the full set of molecules, thus further validating the choice of
the fitting set. Moreover, it also appears that relaxing the condition
on the second non-electrostatic parameter $\beta$ has minor effects
on the overall ability of the method to reproduce experimental energies,
with MAE improving by 0.02-0.04 kcal/mol for two of the sets considered.
In the third set considered (fitg03) an improvement on the quality
of the fit of the order of 0.11 kcal/mol was found. Although not explicitly
parametrized for it, this last set of parameters (fitg03+$\beta$)
is also the one that better reproduces the solvation energy of water
in water, with an excellent agreement with the experimental value
of -6.30 kcal/mol (see Table (\ref{tab:Mean-absolute-error})). Nonetheless,
the relative improvement in the accuracy of the results, obtained
by relaxing the condition on the $\beta$ parameter, is still marginal
compared to the overall performances of the method. 

\begin{table}
\begin{tabular}{|>{\centering}p{1.8cm}|c|c|c|c|c|c|c|c|}
\hline 
\multirow{3}{1.8cm}{$\quad$Name} & \multirow{3}{*}{$\rho^{min}$ (a.u.)} & \multirow{3}{*}{$\rho^{max}$ (a.u.)} &  & \multirow{3}{*}{$\beta$ (GPa)} & $\Delta G^{sol,sccs}$ & MAE  & MAE & MAE \tabularnewline
 &  &  & $\alpha+\gamma$  &  & $\mathrm{H}_{2}\mathrm{O}$ & $\Delta G^{el,g03}$  & $\Delta G^{el,g09}$ & $\Delta G^{sol,exp}$\tabularnewline
 &  &  & (dyn/cm) &  & (kcal/mol) & (kcal/mol) & (kcal/mol) & (kcal/mol)\tabularnewline
\hline 
fitg09 & 0.0001 & 0.0015 & 2.5 & 0.0 & -5.13 & - & 0.27 & 1.53\tabularnewline
fitg09+$\beta$ & 0.0001 & 0.0015 & 11 & -0.08 & -4.90 & - & 0.27 & 1.51\tabularnewline
fitg03 & 0.0001 & 0.0050 & 11.5 & 0.0 & -7.40 & 0.95 & - & 1.31\tabularnewline
fitg03+$\beta$ & 0.0001 & 0.0050 & 50 & -0.35 & -6.29 & 0.95 & - & 1.20\tabularnewline
fitg03' & 0.0003 & 0.0030 & 12 & 0.0 & -7.55 & 0.90 & - & 1.32\tabularnewline
fitg03'+$\beta$ & 0.0003 & 0.0030 & 20 & -0.08 & -7.35 & 0.90 & - & 1.28\tabularnewline
\hline 
\end{tabular}\caption{\label{tab:Mean-absolute-error}Mean absolute error on the total set
of molecules for different sets of parameters. For each set of parameters,
the solvation energy of water in water (experimental value of -6.30
kcal/mol) has also been reported.}
\end{table}

Indeed, reported errors show that the proposed method is not yet able
to be used as a tool to quantitatively predict free energies of solvation.
Nonetheless, when compared with similar computational methods, the
above results are remarkable, since they have been obtained with a
model based on only two independent parameters. In particular, while
the model of Cramer and Truhlar is able to achieve a MAE of the order
of 0.55 kcal/mol for solvation of neutral compounds in aqueous solution
\cite{cramer_accchemres_2008}, it relies on a much higher number
of tunable parameters. Similarly, the use of more tunable methods
to describe the dispersion and repulsion contributions were shown
to improve the agreement with experiment for other IEF-PCM based methods,
with MAE of 0.58 kcal/mol and 1.01 kcal/mol reported for the COSMOtherm
and the MST methods (see Ref. \cite{klamt_accchemres_2009} and references
therein).

\subsubsection{\label{sub:Errors-vs-functional}Errors vs functional groups}

\begin{figure}
\includegraphics[width=1\textwidth]{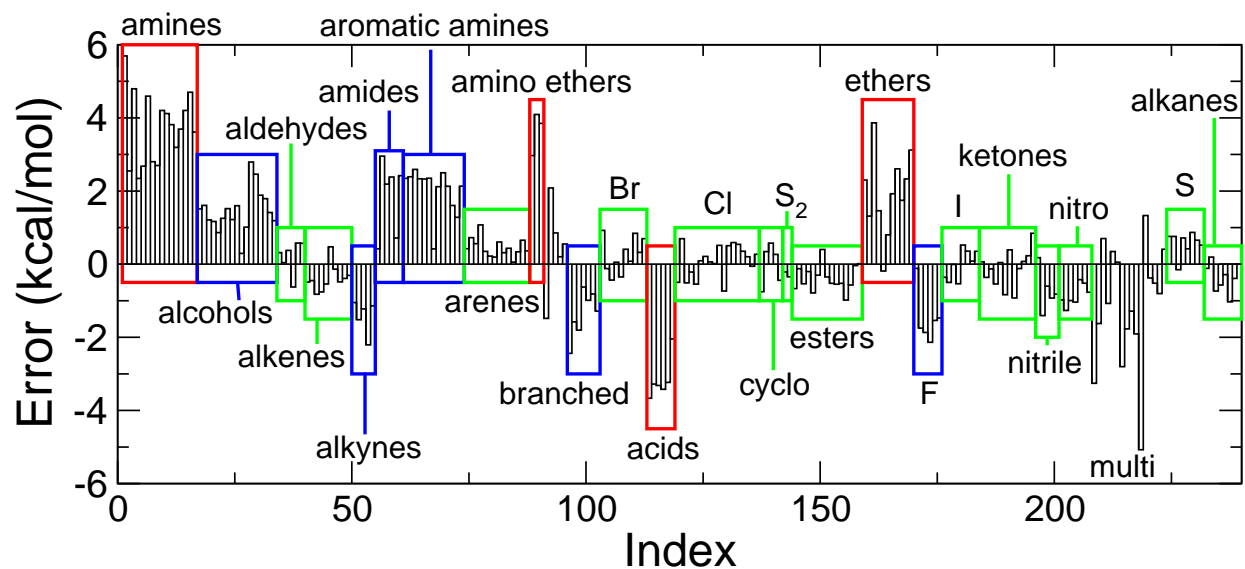}

\caption{\label{fig:Errors-of-Set}Error on solvation energies, computed with
SCCS and the fitg03+$\beta$ set of parameters, for the extended set
of molecules, divided according to their main functional groups (branched
= branched alkanes; acids = carboxylic acids; cyclo = cycloalkanes;
S$_{2}$ = disulfides; multi = multiple halogens). Some compounds
with more than one functional groups have not been explicitly classified.
Colored rectangles have been used to identify the different classes
of compounds, with different colors reflecting the degree of accuracy
of the method: green with maximum errors less than $\pm$1.5 kcal/mol;
blue for maximum errors less than $\pm$3.0 kcal/mol; red for sets
with larger errors (up to $\pm$6.0 kcal/mol) }
\end{figure}

In order to better understand which physical aspects mostly limit
the accuracy of the method, the errors on solvation energies, computed
with the fitg03+$\beta$ parameters have been reported in Figure (\ref{fig:Errors-of-Set})
for all the 240 molecules of the extended set classified according
to their main functional groups. The performances of the method appear
to be very good for a large set of organic molecules, including linear
and cyclic alkanes, alkenes, arenes, aldehydes, ketones, and esters.
With the noticeable exception of fluorine, etheroatoms and halogens
do not affect the accuracy of the results, and compounds containing
chlorine, bromine, iodine, mono- and di- sulfides, nitro and nitrile
groups show reasonable errors (the largest deviations being less than
1.0 kcal/mol), well within the accuracy of the best implicit solvation
computational methods. Considering only the above classes of compounds,
a moderate MAE of 0.46 kcal/mol is obtained. 

Instead, the performance of the method is poor for two specific types
of functional groups: amines and carboxylic acids, that show positive
(amines) and negative (acids) errors on solvation energies of up to
4-5 kcal/mol. The acid and basic natures of these two classes of compounds
is probably the cause of large discrepancies between computed and
experimental results. In particular, SCCS completely neglects the
possibility that different species of the same solute could be in
equilibrium in solution, corresponding to more or less protonation/deprotonation.
Thermodynamically meaningful solvation energies should instead be
computed as a weighted average of the solvation energy of the neutral
species and that of the protonated/deprotonated species. A refinement
of the fit to include the effects of acid-base equilibria will be
the subject of further developments. 

In addition to the acid/basic compounds, poor results are obtained
also for ethers, alcohols and amides. Whether this is due to the lack
of a correct description of the hydrogen bonding environment with
a continuum model is still under study. Nonetheless, the better performance
of alcohols with respect to ethers seem to suggest that hydrogen bonding
should not be the only reason for such discrepancies. Fluorine compounds,
and in particular multifluorinated alcohols and ethers, also show
poor results compared to experiment. This effect could be again due
to the lack of explicit hydrogen bonding in SCCS, that should affect
fluorinated compounds more than other halogenated molecules. Moreover,
the discrepancies in the computed results could be due to the different
numerical treatment given to molecules containing fluorine, where
higher density and wavefunction cutoffs were used in order to compensate
for the hardness of the fluorine pseudopotentials. Eventually, alkynes
and branched alkanes show somewhat poor agreement with experiment,
probably due to the unideal description of the cavitation and repulsion
potentials.

\subsubsection{\label{sub:Molecular-Dynamics-simulations}Molecular dynamics simulations}

\begin{figure}
\includegraphics[width=0.9\textwidth]{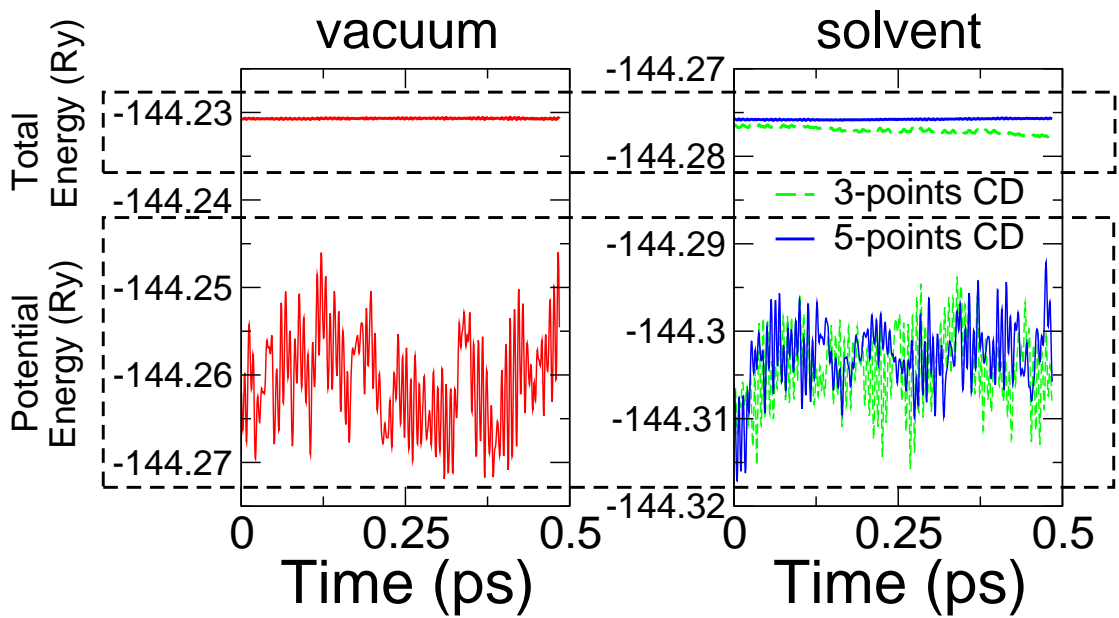}

\caption{\label{fig:Total-Energy-Conservation}Total and potential energies
of a water tetramer along Born-Oppeneimer NVE molecular dynamics trajectories,
in vacuum (left panel) and in a continuum acqueous solution (right
panel). The same accuracy on total energy conservation is found for
the simulation in vacuum and in solution, when a high-order (e.g.
5-points, blue curves) central differences (CD) scheme is exploited
to compute in real space the gradient of the dielectric function.
When the simpler 3-points CD scheme (green curves) is adopted, a small
systematic error on the interatomic forces results in poor energy
conservation along the MD simulation.}

\end{figure}

In order to validate the accuracy of the proposed method for MD simulations,
a prototypical system composed by a cyclic tetramer of deuterated
water molecules has been studied, similarly to what was done in Ref.
\cite{scherlis_jcp_2006}. As reported in the numerical details subsection
\ref{sub:Numerical-Details}, convergence of constant energy simulations
in vacuum, as reflected by total energy conservation, required to
sensibly tighten some of the key parameters of the simulation. When
the same parameters of the calculations in vacuum were adopted for
a simulation in the presence of continuum aqueous solution, a sensible
worsening of the accuracy of the method was found. This behaviour
is due to a small inaccuracy in the calculation of the forces, that
has no significant effects on the geometry relaxations of the tested
molecules. Such an error could be traced back to be related to the
use of a simplified 3-points central difference schemes in the evaluation
of the gradient of the dielectric that appears in the definition of
the polarization charges (see, e.g., Eqs. (\ref{eq:rhopol_final})
and (\ref{eq:rhoiter_def})). Several alternative and more advanced
schemes can be adopted for a finite difference calculation of such
a gradient, including higher order central differences or smooth noise-robust
differentiators \cite{snrd}. By exploiting a number of points in
the discretized space larger than three, these schemes can produce
an unphysical non-vanishing gradient in the flat regions of space
close to the solvent-vacuum interface. Nonetheless, the more accurate
description of the gradient in the interfacial region produces interatomic
forces more accurate than the ones obtained by the simplest 3-points
central difference method, thus resulting in total energy conservation
of the same quality of the simulation in vacuum (compare left and
right panel of Figure (\ref{fig:Total-Energy-Conservation})). 

\section{Conclusions}

A revised self-consistent continuum solvation model has been presented
that overcomes most of the limitations of previous approaches and
includes, in a simplified way, additional non-electrostatic contributions
to solvation free energies via a combined use of the concepts of quantum
volume and quantum surface \cite{cococcioni_prl_2005} for isolated
fragments. The model has been implemented using a novel iterative
approach that is robust, efficient, fully parallel and convenient
to implement in electronic-structure codes, both for isolated or periodic
geometries independently of the charge density representation or the
Bravais lattice chosen. Parametrization of the method to reproduce
the experimental results has been performed, and the overall performance
over an extended set of solvation free energies in water compares
favorably with the results from reference theoretical methods of similar
physical background and more parameter intensive. The static dielectric
constant of the solvent and one parameter allow to fit the electrostatic
energy provided by the PCM model with a mean absolute error of 0.3
kcal/mol on a set of 240 neutral solutes, and two parameters allow
to fit experimental solvation energies on the same set with a mean
absolute error of 1.3 kcal/mol. A detailed analysis of these results,
broken down along different classes of chemical compounds, shows that
most molecules display even closer agreement, whereby larger errors
are mostly limited to self-dissociating species and strong hydrogen-bond
forming compounds. Last, a careful analysis of the effects of numerical
parameters on the predictions of the method have been presented, to
make present results fully reproducible.
\begin{acknowledgments}
This work was carried our as part of the MIT Energy Initiative, with
the support of Robert Bosh LLC. The authors would like to thank D.
Scherlis, J.-L. Fattebert, C.-K. Skylaris, C.-H. Park and J. Tomasi
for a careful reading of the manuscript and their suggestions. We
are particularly grateful to B. Mennucci for useful discussions and
her help with the continuum solvation literature. 
\end{acknowledgments}

\section*{APPENDIX A: \label{sub:Effect-of-numerical}Effect of numerical parameters}

In an effort of making the present results reproducible and to study
the accuracy of the model proposed, a careful analysis of the effect
of all other numerical and computational parameters on solvation energies
is mandatory. In fact, only a few of these have been discussed and
shown to have a negligible effect on the final results. This is the
case of the finite-difference parameter $\Delta$ that appears in
the definition of the cavitation energy, Eq. (\ref{eq:cavitation_potential}),
which was found to have negligible effects on the computed energies
\cite{scherlis_jcp_2006}. For this reason, a value of $\Delta=0.0001$
a.u. was used throughout the simulations. Similarly, as discussed
in Section \ref{sec:Iterative-vs-Multigrid}, the parameters entering
into the iterative algorithm to compute the polarization charges show
no significant impact on the final results. 

The main parameter that enters into an ab-initio calculation is the
size of the basis set, that in the plane-wave codes for which the
model was developed corresponds to the kinetic energy cutoffs of the
plane waves used to represent the wavefunctions and electronic density.
The density cutoff, in particular, corresponds to the grid size in
real space used in all the calculations of the polarization density
(see Section (\ref{sec:Iterative-vs-Multigrid})) and, as such, it
represents a crucial numerical parameter of the model. Since the electronic
density is given by the square of the wavefunction, a cutoff on the
density four times larger than the one on the wavefunction is usually
adopted. Nonetheless, the use of ultrasoft pseudopotential, while
allowing to lower the plane-wave cutoffs for the description of the
wavefunction, requires a much higher ratio between the two cutoffs,
with density cutoffs 8 to 10 times larger than the ones on wavefunctions. 

\begin{figure}
\includegraphics[width=0.9\textwidth]{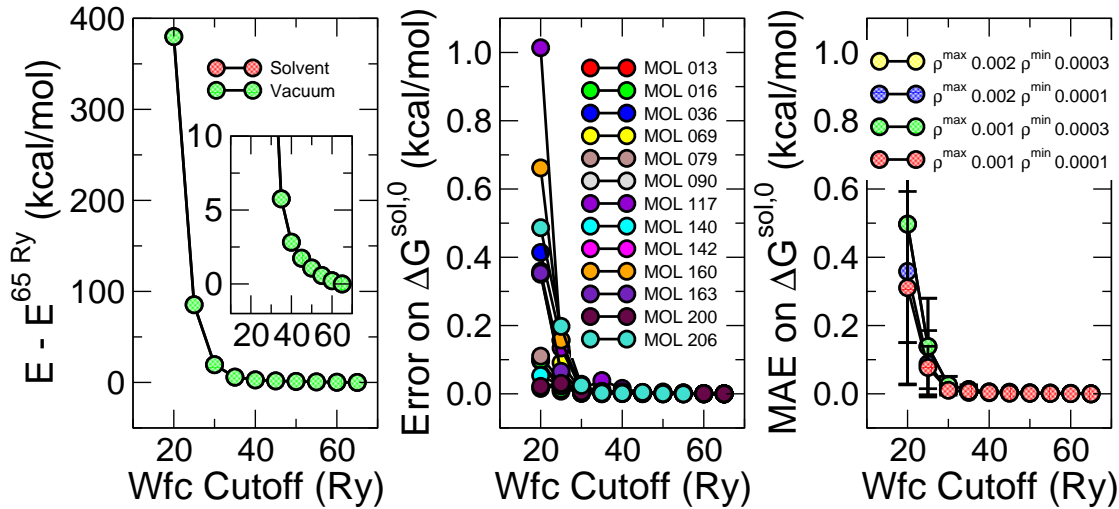}

\caption{\label{fig:dgsol-vs-wcut}Convergence of the calculations with respect
to wavefunction (wfc) cutoff. Left: convergence of total energy of
molecule 013 of set (piperazine) in vacuum and in solution (solvent
parameters: $\rho^{min}=0.0001$ a.u., $\rho^{max}=0.001$ a.u.).
Note that the two curves are on top of each other on the scale of
the graph. In the inset, a closer view on the convergence of the error
for large wavefunction cutoffs is reported. Center: convergence of
the absolute error on solvation free energy $\Delta G^{sol,0}$ for
the thirteen molecules of the fitting set (solvent parameters: $\rho^{min}=0.0001$
a.u., $\rho^{max}=0.001$ a.u.). Right: convergence of the mean absolute
error on solvation free energies $\Delta G^{sol,0}$ of the fitting
set for different values of solvent parameters. Error bars represent
one standard deviation from the average.}
\end{figure}

Convergence of the electrostatic solvation free energy with wavefuntion
cutoffs for the thirteen molecules of the fitting set is reported
in Figure (\ref{fig:dgsol-vs-wcut}). Results show a reasonably fast
convergence with respect to wavefunction cutoff: while variations
on the computed energies are of the order of the kcal/mol even for
the larger cutoffs considered, solvation free energies are well converged
for wavefunction cutoffs of the order of 30 Ry. At lower cutoffs a
small subset of molecules present errors on solvation energies of
the order of 0.5-1.0 kcal/mol: since the molecules on this subset
are the ones containing oxygen, it seems that the error is mostly
related to the use of a pseudopotential for oxygen that is harder
than the ones used for the other atomic types. The fact that changing
the solvent parameters has a negligible effect on the convergence
of the results also supports the conclusion that the solvation calculation
is independent on the choice of the wavefunction cutoffs. 

\begin{figure}
\includegraphics[width=0.9\textwidth]{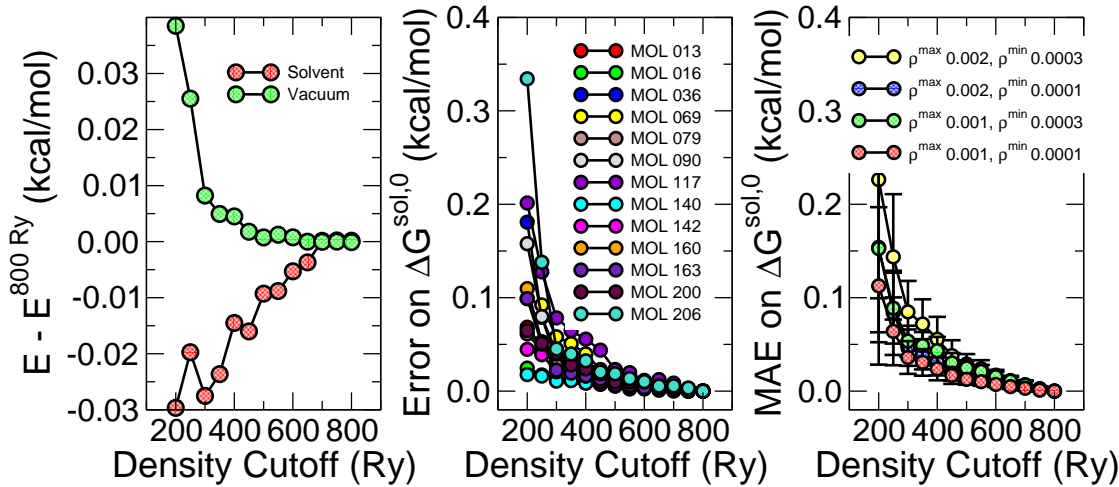}\caption{\label{fig:dgsol-vs-ecut}Convergence of the calculations with respect
to density cutoff. Left: convergence of total energy of molecule 013
of set (piperazine) in vacuum and in solution (solvent parameters:
$\rho^{min}=0.0001$ a.u., $\rho^{max}=0.001$ a.u.). Center: convergence
of the absolute error on solvation free energy $\Delta G^{sol,0}$
for the thirteen molecules of the fitting set (solvent parameters:
$\rho^{min}=0.0001$ a.u., $\rho^{max}=0.001$ a.u.). Right: convergence
of the mean absolute error on solvation free energies $\Delta G^{sol,0}$
of the fitting set for different values of solvent parameters. Error
bars represent one standard deviation from the average.}
\end{figure}

Different results are obtained for the convergence with respect to
the density cutoff, as shown in Figure (\ref{fig:dgsol-vs-ecut}).
Convergence is slightly slower than with respect to wavefunction cutoffs,
with errors of the order of 0.2-0.3 kcal/mol for the normal range
of this parameter (i.e. between 300 and 400 Ry). This residual error
appear to be mostly related in the first place to the sharpness of
the polarization charge and potentials. Moreover, the numerical details
of the method could also be crucial, as reflected by the fact that
the polarization charge density is not neutral. In particular, the
calculation of $\mathbf{\nabla}\ln\epsilon\left[\rho^{elec}\left(\mathbf{r}\right)\right]$
that enters into Eq. (\ref{eq:rhoiter_def}) is performed using a
simplified approach, that involves a centered three-point finite difference
algorithm in real space. Nonetheless, for the range of solvent parameters
investigated, the error is generally less than 0.1 kcal/mol, thus
well within the typical accuracies of conventional continuum solvation
models. 

\begin{figure}
\includegraphics[width=0.9\textwidth]{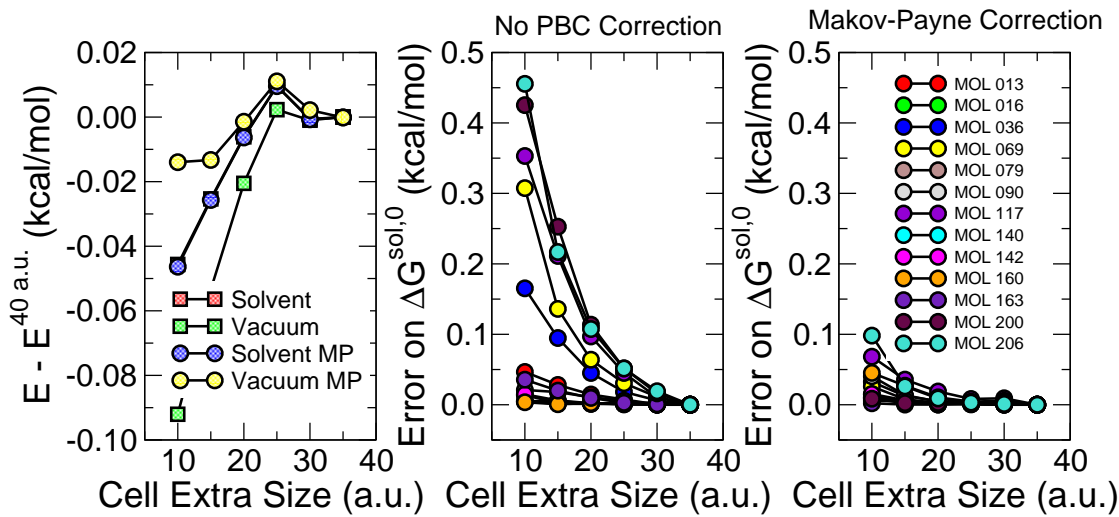}\caption{\label{fig:dgsol-vs-cellsize}Convergence of the calculations with
respect to cell extra size, where effective cell size is defined as
the maximum dimension of the molecule plus cell extra size. Left:
convergence of total energy of molecule 013 of set (piperazine) in
vacuum and in solution, with no periodic boundary corrections (squares)
and with Makov-Payne correction (circles) applied. Center: convergence
of the absolute error on solvation free energy $\Delta G^{sol,0}$
for the thirteen molecules of the fitting set with no periodic boundary
corrections. Right: convergence of the absolute error on solvation
free energy $\Delta G^{sol,0}$ for the thirteen molecules of the
fitting set with Makov-Payne periodic boundary correction. }
\end{figure}

In Figure (\ref{fig:dgsol-vs-cellsize}) the cell-size dependence
of the electrostatic contribution to solvation free energies is reported,
for the thirteen molecules of the fitting set, with and without corrections
for periodic-boundary conditions. As already mentioned, cell sizes
were chosen to be equal to the maximum size of the molecule optimized
in vacuum, plus an adjustable amount of 10 to 40 a.u. (denoted cell
extra size in Figure (\ref{fig:dgsol-vs-cellsize})). Since calculations
are performed for a solvent with a high dielectric constant, screening
of solute charges makes the effects of periodic-boundary conditions
relatively small. When including corrections for periodic-boundary
conditions, such as in the Makov-Payne scheme \cite{makov_prb_1995},
no significant variations were found for solvation energies calculated
from Eq. (\ref{eq:dg_electrostatic_unrelaxed}), while large changes
of the energy in vacuum appear. Thus, the error is mostly related
to the accuracy of the calculation in vacuum. Moreover, the overall
error was found to be lower than 0.1 kcal/mol for most of the molecules
in the fitting set, while larger errors up to 0.5 kcal/mol were found
for the molecules with larger dipole moments. When a correction for
periodic images is explicitly accounted for, convergence is quite
fast even for the smallest cell sizes considered. More advanced and
accurate periodic boundary correction schemes, such as the density-countercharge
correction (DCC) \cite{dabo_prb_2008}, can be easily extended to
calculations with the SCCS, by explicitly considering the polarization
charge density.

\begin{figure}
\includegraphics[width=0.8\textwidth]{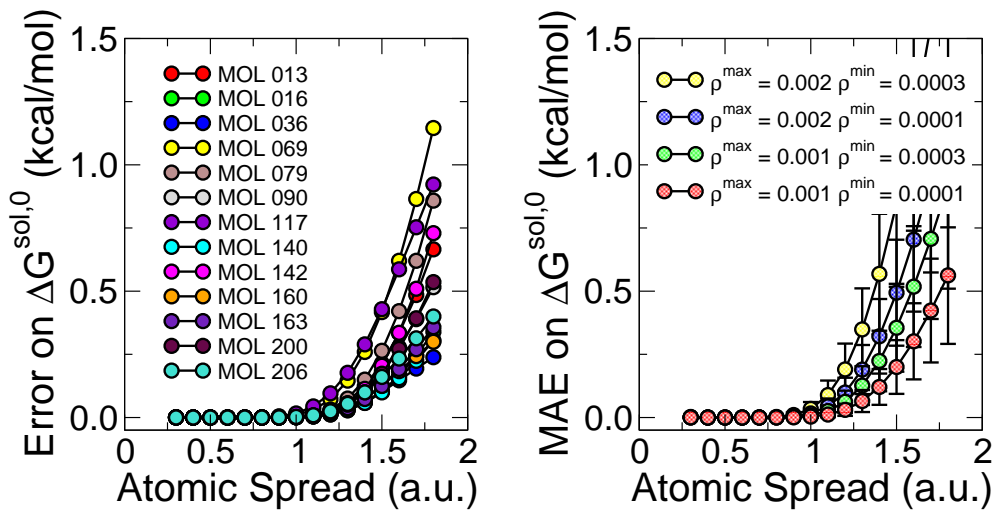}\caption{\label{fig:dgsol-vs-atomicspread}Convergence of the calculations
with respect to atomic spreads. Center: convergence of the absolute
error on solvation free energy $\Delta G^{sol,0}$ for the thirteen
molecules of the fitting set (solvent parameters: $\rho^{min}=0.0001$
a.u., $\rho^{max}=0.001$ a.u.). Right: convergence of the mean absolute
error on solvation free energies $\Delta G^{sol,0}$ of the fitting
set for different values of the solvent parameters. Error bars represent
one standard deviation from the average.}
\end{figure}

Last, the effect on solvation free energies of the fictitious atomic
density used to compute the electrostatic field of nuclei and core
electrons has to be taken into account (see Figure (\ref{fig:dgsol-vs-atomicspread})).
In order to study such an effect, we assumed the nuclei to be described
by Gaussians of fixed spread $\sigma$ that do not depend on the atomic
type. This is a typical approximation in plane-wave codes, and the
correct electrostatic is usually recovered in the total energy by
including Ewald complementary-error-function electrostatic terms.
Note that more advanced approaches can be adopted, e.g. by defining
Gaussian spreads to depend on pseudopotential radii, or by explicitly
accounting for the core electronic density. 

As explained in Section (\ref{sec:Choice-of-dielectric}), in order
for the energy of the solvated system to be well defined, the fictitious
ionic charge density should be entirely included in the solvent exclusion
region. Such a region is directly related to the parameters $\rho^{min}$
and $\rho^{max}$ used in the definition of the dielectric constant.
In particular, a high value of $\rho^{max}$ implies the presence
of dielectric contributions close to the nuclei. As a result, for
higher values of $\rho^{max}$, the effect of the ionic density becomes
sensitive at lower values of $\sigma$. Nonetheless, in the whole
range of density thresholds considered we always obtain well converged
results for Gaussian spreads in the window $0.3\textrm{ a.u.}<\sigma<1.0$
a.u..

\end{document}